\newif\ifsubmode
\submodefalse

\newif\ifprintfig
\printfigtrue

\newif\ifemulate
\emulatetrue

\ifsubmode
  \documentclass[12pt,preprint]{aastex}
  \received{}
  \accepted{}
  \journalid{}{}
  \articleid{}{}
\else
   \documentclass{emulateapj}
   \submitted{{\it To be submitted for publication in ApJ}}
\fi

\usepackage{amsmath}

\newcommand{\kms}{\,km~s$^{-1}$}
\newcommand{\Msun}{\mbox{\,$M_{\odot}$}}

\def\lesssim{\mathrel{\hbox{\rlap{\hbox{\lower4pt\hbox{$\sim$}}}\hbox{$<$}}}}
\def\gtrsim{\mathrel{\hbox{\rlap{\hbox{\lower4pt\hbox{$\sim$}}}\hbox{$>$}}}}
\def\spose#1{\hbox to 0pt{#1\hss}}
\def\simlt{\mathrel{\spose{\lower 3pt\hbox{$\mathchar"218$}}
     \raise 2.0pt\hbox{$\mathchar"13C$}}}
\def\simgt{\mathrel{\spose{\lower 3pt\hbox{$\mathchar"218$}}
     \raise 2.0pt\hbox{$\mathchar"13E$}}}

\slugcomment{Draft Version \today}

\shorttitle{Diffuse Ionized Gas in NGC 5775}
\shortauthors{Boettcher, Gallagher, and Zweibel}

\begin{document}

\title{A Dynamical Study of Extraplanar Diffuse Ionized Gas in NGC 5775$^{*}$}
\thanks{$^{*}$Based on observations made with the Southern African Large
  Telescope (SALT) under programs 2015-1-SCI-023 and 2016-2-SCI-029 (PI:
  E. Boettcher).}

\author{Erin Boettcher$^{1,2}$, J. S. Gallagher III$^{1}$, Ellen
  G. Zweibel$^{1,3}$}

\affiliation{$^{1}$Department of Astronomy, University of Wisconsin - Madison,
  475 North Charter Street, Madison, WI 53706, USA}
\affiliation{$^{2}$Department of Astronomy \& Astrophysics, University of
  Chicago, 5640 S. Ellis Avenue, Chicago, IL 60637, USA; \texttt{eboettcher@astro.uchicago.edu}}  
\affiliation{$^{3}$Department of Physics,
  University of Wisconsin - Madison, 1150 University Avenue, Madison, WI
  53706, USA} 


\ifsubmode\else
  \ifemulate\else
     \clearpage
  \fi
\fi


\ifsubmode\else
  \ifemulate\else
     \baselineskip=14pt
  \fi
\fi

\begin{abstract} 

  The structure and kinematics of gaseous, disk-halo interfaces are imprinted
  with the processes that transfer mass, metals, and energy between galactic
  disks and their environments. We study the extraplanar diffuse ionized gas
  (eDIG) layer in the interacting, star-forming galaxy NGC 5775 to better
  understand the consequences of star-formation feedback on the dynamical
  state of the thick-disk interstellar medium (ISM). Combining emission-line
  spectroscopy from the Robert Stobie Spectrograph on the Southern African
  Large Telescope with radio continuum observations from Continuum Halos in
  Nearby Galaxies - an EVLA Survey, we ask whether thermal, turbulent,
  magnetic field, and cosmic-ray pressure gradients can stably support the
  eDIG layer in dynamical equilibrium. This model fails to reproduce the
  observed exponential electron scale heights of the eDIG thick disk and halo
  on the northeast ($h_{z,e} = 0.6, 7.5$ kpc) and southwest ($h_{z,e} = 0.8,
  3.6$ kpc) sides of the galaxy at $R < 11$ kpc. We report the first
  definitive detection of an increasing eDIG velocity dispersion as a function
  of height above the disk. Blueshifted gas along the minor axis at
  large distances from the midplane hints at a disk-halo circulation and/or
  ram pressure effects caused by the ongoing interaction with NGC 5774. This
  work motivates further integral field unit and/or Fabry-Perot spectroscopy
  of galaxies with a range of star-formation rates to develop a
  spatially-resolved understanding of the role of star-formation feedback in
  shaping the kinematics of the disk-halo interface.
  
\end{abstract}


\section{Introduction}
\label{chap3:sec:intro}

In Milky Way-type galaxies, the gaseous, disk-halo interface is shaped by
processes that regulate the growth of galaxies at low redshift. These
processes include star-formation feedback, which drives galactic fountains
\citep{Shapiro1976, Bregman1980} and galactic winds
\citep[e.g.,][]{Veilleux2005}, as well as cold gas accretion necessary to
sustain star formation over cosmic lifetimes \citep[e.g.,][]{Larson1980}. Well
known galaxy scaling relationships, including the stellar-mass - halo-mass and
mass-metallicity relations \citep[e.g.,][]{Silk2003, Tremonti2004}, suggest
that mass, metals, and energy cycle through the disk-halo interface in ways
that have important implications for galaxy evolution.

These observations motivate a careful study of the vertical structure,
support, and kinematics of the gaseous, disk-halo interfaces of nearby
galaxies. A long-standing challenge to our understanding of the dynamical
state of the thick-disk interstellar medium (ISM) is the presence of
extraplanar diffuse ionized gas (eDIG) layers in the Milky Way and other
star-forming galaxies. Vertically-extended, photoionized, diffuse ($T \sim
10^{4}$ K, $\langle n_{e,0} \rangle \sim 0.1$ cm$^{-3}$) gas is observed in
galaxies with star-formation rates comparable to or exceeding that of the
Galaxy \citep{Lehnert1995, Rossa2003a}. A particularly puzzling feature of
eDIG layers is that their observed exponential electron scale heights
($h_{z,e} \sim 1$ kpc) exceed their thermal scale heights ($h_{z,e,th} \sim
0.2$ kpc) by factors of a few in the Milky Way \citep{Haffner1999} and nearby,
edge-on disk galaxies \citep[e.g.,][]{Rand1997, Collins2001}. As yet, the
large scale heights of eDIG layers have defied full explanation with ballistic
\citep[e.g.,][]{Collins2002} and (magneto)hydrodynamic models
\citep[e.g.,][]{Hill2012}.

A simple question that can be asked is whether there is sufficient vertical
pressure to stably support eDIG layers in dynamical equilibrium if turbulent,
magnetic field, and cosmic-ray pressure gradients are invoked in addition to
the thermal pressure gradient. While dynamical equilibrium has been considered
as a viable model for the vertically-extended ISM in the Milky Way
\citep{Boulares1990, Fletcher2001}, this question has only recently been
considered in detail for eDIG layers in external
galaxies. \citet{Boettcher2016} demonstrated that thermal and nonthermal
pressure gradients are only capable of stably supporting the eDIG layer in NGC
891 at large galactocentric radii ($R \ge 8$ kpc). However, application of
such analysis to other galaxies is desired, including those with differing
star-formation rates and interaction histories. In the absence of dynamical
equilibrium, it is then apt to ask whether there is kinematic evidence for a
nonequilibrium process such as a galactic fountain, wind, or accretion flow.

Here, we perform a dynamical study of the eDIG layer in the nearby, edge-on
disk galaxy NGC 5775 (SBc? sp; \citealt{deVaucouleurs1991}). This
well-studied, star-forming galaxy is viewed at an inclination angle of $i =
86^{\circ}$ \citep{Irwin1994}, and we adopt a distance of $D = 28.9$ Mpc ($1''$
= 0.14 kpc; \citealt{Irwin2012}). NGC 5775 has an \textit{IRAS} far-infrared
surface brightness of $L_{FIR}/D_{25}^{2} = 8.4 \times 10^{40}$ erg s$^{-1}$
kpc$^{-2}$ that is consistent with the more quiescent of starburst systems
\citep{Collins2000, Rossa2003a}. One of its most noteable features is its
interaction with companion galaxy NGC 5774; there is evidence that this barred
spiral galaxy is donating neutral hydrogen gas to NGC 5775 via an HI bridge
characterized by \citet{Irwin1994}, enhancing the star-formation rate of the
latter galaxy. NGC 5775 has a spatially-extended, multiphase gaseous halo
whose structure and kinematics shed light on the disk-halo connection in
star-forming systems \citep[e.g.,][]{Irwin1994, Rand2000, Tullmann2006,
  Li2008}.

The structure, ionization, and kinematics of the eDIG layer in NGC 5775 have
been the subject of extensive study. Narrow-band imaging reveals a highly
filamentary eDIG layer superimposed on a diffuse background
\citep{Collins2000}. This galaxy has one of the most vertically-extended
emission-line halos known; \citet{Rand2000} detected emission to $z = 15$ kpc,
corrected for our choice of $D$. The evolution of the emission-line ratios
with height above the disk suggests the presence of a supplementary source of
heating and/or ionization in addition to photoionization, such as shocks or
turbulent mixing layers \citep[e.g.,][]{Tullmann2000, Collins2001,
  Otte2002}. The interplay, if any, between dynamical processes in the eDIG
and supplementary sources of heating and ionization remains an open
question. \citet{Heald2006a} probe the kinematics of the eDIG layer using
Fabry-Perot spectroscopy, revealing a rotational velocity lag in the
extraplanar gas within a few kiloparsecs of the disk.

Consideration of the role of extraplanar magnetic fields and cosmic rays in
the dynamics, heating, and/or ionization of the eDIG in NGC 5775 is motivated
by the detection of an extensive ($z \sim 12 - 17$ kpc, corrected for $D$)
radio continuum halo \citep{Duric1998}. More recently, the synchrotron halo of
NGC 5775 has been studied by Continuum Halos in Nearby Galaxies - an EVLA
Survey (CHANG-ES). A goal of this program is to characterize the nonthermal
halos of 35 nearby, edge-on disk galaxies at 1.5 and 6 GHz with the Karl
G. Jansky Very Large Array \citep{Irwin2012, Wiegert2015}. A Faraday rotation
study by \citet{Soida2011} examined the large-scale structure of the magnetic
field in NGC 5775, revealing a plane-parallel field close to the disk and an
increasingly X-shaped field in the halo. \citet{Collins2000} present evidence
for a spatial correlation between H$\alpha$ filaments, HI shells, and radio
continuum ``spurs'' suggestive of a galactic chimney model. These correlations
motivate a better understanding of how gas, magnetic fields, and cosmic rays
are transported by disk-halo flows and whether or not they are able to achieve
a mutual equilibrium state.

This paper is presented as follows. In \S\ref{chap3:sec:obs}, we describe the
collection, reduction, and analysis of optical and near-ultraviolet (NUV)
longslit spectroscopy obtained with the Robert Stobie Spectrograph (RSS) on
the Southern African Large Telescope (SALT). In \S\ref{chap3:sec:Ha_int}, we
characterize the eDIG density distribution, and we discuss the emission-line
ratios and their implications for the physical conditions in the eDIG layer in
\S\ref{chap3:sec:eDIG_prop}. We describe the radial velocities and velocity
dispersions in \S\ref{chap3:sec:eDIG_kin}. In \S\ref{chap3:sec:dyn_eq}, we
develop and test a dynamical equilibrium model of the eDIG layer and consider
whether it is robust against the Parker instability. We discuss our results in
the context of the literature in \S\ref{chap3:sec:disc} and conclude with
motivation for future observations in \S\ref{chap3:sec:summ}.

\section{Observations}
\label{chap3:sec:obs}

\subsection{Data Collection}
\label{chap3:sec:data_coll}

\begin{figure}[h]
  \epsscale{1.2}\plotone{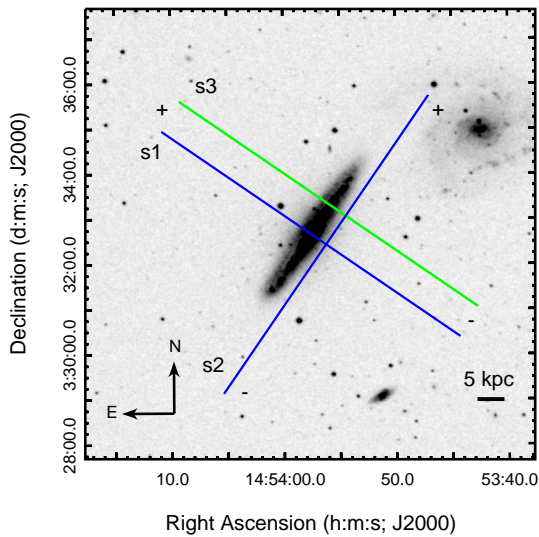}
  \caption{SALT-RSS longslits
  overplotted on a POSS2/UKSTU Red image of NGC 5775 from the
  Digitized Sky Survey (Second
  Generation; http://stdatu.stsci.edu/cgi-bin/dss\textunderscore
  form). We obtained optical observations with moderate spectral
  resolution at the locations of the blue slits and NUV
  ([OII]$\lambda\lambda$3726, 3729; moderate resolution) and optical
  (low resolution) observations at the location of the green slit. The
  plus and minus signs indicate the sign conventions applied to the
  vertical ($z$) and radial ($R$) coordinates.}
  \label{chap3:fig1}
\end{figure}

We used the optical and NUV capabilities of RSS \citep{Burgh2003,
  Kobulnicky2003} in longslit mode on SALT \citep{Buckley2006}. We placed one
slit on the minor axis (PA = $55.7^{\circ}$, s1) and a second slit parallel to
the major axis and offset by $\Delta z = 3.5$ kpc on the southern side of the
galaxy (PA = $145.7^{\circ}$, s2; see Fig. \ref{chap3:fig1}). We used a
$1.25''$ width for the $8'$ longslits and the pg2300 grating at an angle of
$48.875^{\circ}$. This produced a dispersion of $0.26\ \text{\AA}\
\text{pixel}^{-1}$, a spectral resolution of $\texttt{R} = 4830$ ($\sigma =
26$ \kms) at H$\alpha$, and wavelength coverage from $6100\ \text{\AA}$ to
$6900\ \text{\AA}$, including the [NII]$\lambda\lambda$6548, 6583, H$\alpha$,
and [SII]$\lambda\lambda$6717, 6731 emission lines. We obtained these data
between 2017 February 23 and 2017 March 04.

We also placed a slit perpendicular to the disk at $R = 6.5$ kpc on the
northwest side of the galaxy (PA = $55.7^{\circ}$, s3; see Fig.
\ref{chap3:fig1}). This slit falls along a bright filament detected in
H$\alpha$ imaging by \citet{Collins2000}. We obtained two sets of observations
at this location: moderate spectral resolution observations of the
[OII]$\lambda\lambda$3726, 3729 emission-line doublet (s3uv), and low spectral
resolution observations with wavelength coverage from
[OII]$\lambda\lambda$3726, 3729 to [NII]$\lambda\lambda$6548, 6583 (s3o). The
[OII] doublet is a powerful eDIG tracer because it is comparably bright to
H$\alpha$ \citep[e.g.,][]{Otte2002} and is found in the NUV regime where the
terrestrial sky foreground is minimal compared to the optical.

For the s3uv observations, a $1.5''$ slit width and the pg3000 grating at an
angle of $34.25^{\circ}$ yielded a dispersion of $0.24\ \text{\AA}\
\text{pixel}^{-1}$ and a spectral resolution of $\texttt{R} = 2400$ ($\sigma =
53$ \kms) at [OII]. For s3o, the pg0900 grating at an angle of
$13.625^{\circ}$ produced a dispersion of $0.97\ \text{\AA}\
\text{pixel}^{-1}$ and a spectral resolution of $\texttt{R} = 620$ ($\sigma =
205$ \kms) at [OII] and $\texttt{R} = 1140$ ($\sigma = 112$ \kms) at
H$\alpha$. We obtained these observations between 2015 June 17 and 2015 August
06. For all data, we used $2 \times 2$ on-chip binning to yield a plate scale
of $0.25''\ \text{pixel}^{-1}$. We note the coordinates, position angles, and
exposure times in Tab. \ref{chap3:tab1}.

\begin{deluxetable}{ccccc}
\tabletypesize{\scriptsize}
\tablecolumns{5}
\tablewidth{0pt}
\tablecaption{NGC 5775 Observing Summary}
\tablehead{ 
\colhead{Slit} &
\colhead{R.A. \tablenotemark{a}} &
\colhead{Decl. \tablenotemark{a}} &
\colhead{P.A. \tablenotemark{b}} &
\colhead{$t_{exp}$}\\
\colhead{Label} &
\colhead{(J2000)} &
\colhead{(J2000)} &
\colhead{(deg)} &
\colhead{(s)}
}
\startdata
s1 & 14 53 57.6 & +03 32 40 & 55.7 & $5 \times 1600$ \\
s2 & 14 53 56.2 & +03 32 26 & 145.7 & $4 \times 1615$ \\
s3uv & 14 53 56.0 & +03 33 20 & 55.7 & $3 \times 1670$ \\
s3o & 14 53 56.0 & +03 33 20 & 55.7 & $2 \times 1660$
\enddata
\tablenotetext{a}{The R.A. and decl. at the center of the slit. R.A. is
  measured in hours, minutes, and seconds; decl. is measured in degrees,
  arcminutes, and arcseconds.}  \tablenotetext{b}{The position angle measured
  from north to east.}
\label{chap3:tab1}
\end{deluxetable}

\subsection{Data Reduction}
\label{chap3:sec:data_red}

We performed the data reduction using a combination of the PySALT science
pipeline\footnote[1]{\url{http://pysalt.salt.ac.za/}} and IRAF
software\footnote[2]{IRAF is distributed by the National Optical Astronomy
  Observatories, which are operated by the Association of Universities for
  Research in Astronomy, Inc., under cooperative agreement with the National
  Science Foundation.}. We utilized PySALT \citep{Crawford2010} to apply the
bias, gain, and cross-talk corrections as well as for the image preparation
and mosaicking. Using the IRAF \texttt{L.A.Cosmic} package
\citep{vanDokkum2001}, we removed cosmic rays from the images. With the IRAF
tasks \texttt{noao.twodspec.longslit.identify}, \texttt{reidentify},
\texttt{fitcoords}, and \texttt{transform}, we determined the dispersion
solution using Ar comparison lamp spectra for s1 and s2. ThAr and Ar
comparison lamp spectra were used for the low and moderate spectral resolution
observations of s3.

In s1 and s2, there is a count background of currently unknown
origin that varies along both the spatial and spectral
axes. The presence of this background - identified as elevated counts
above the bias - on both the illuminated and non-illuminated portions
of the CCD chip confirm its origin as separate from the sky
continuum. It contributes as much as 85\% of the sky continuum at the
blue end of the spectral range, but less than 20\% of the sky
continuum near the emission lines of interest. Due to the
difficulty in separating the contributions from this background, the
sky, and the galaxy, we fit and remove all three contributions to the
continuum simultaneously. This is done by modeling the continuum pixel
row by row in the spectral direction in the two-dimensional spectrum
using a fifth-order Chebyshev function in the IRAF
task \texttt{noao.twodspec.longslit.background}. We assume
that the unidentified background, sky, and galaxy continuum
all contribute Poisson noise in the error estimation.

Our approach to sky-line subtraction differed depending on the slit
position. For s1 and s3, we created median sky spectra from the ends of the
slits that were free from galaxy emission. Before performing the sky
subtraction, we scaled the sky spectrum by a multiplicative factor that we
allowed to vary in the spatial dimension based on the relative strength of the
sky lines across the slit. For s2, we did not know beforehand what fraction of
the slit would be filled with galaxy emission. We thus obtained a single
separate sky observation immediately following the first object observation
along a telescope track that replicated the object track, exposure time, and
instrument setup. Based on the relative intensities of the sky lines in the
object and sky frames, we scaled the sky observation by a multiplicative
factor before subtracting it from the object observations. Due to the count
background mentioned previously, we fit and removed the continuum from both
the sky and object frames before performing the sky-line subtraction.

We stacked the sky-subtracted science frames by aligning them in the spatial
dimension based on the location of the peak H$\alpha$ or [OII] emission. We
then used the IRAF task \texttt{images.immatch.imcombine} to perform a median
combine of the images with median scaling and weighting. To extract individual
spectra, we performed a median combine over an aperture of 11 pixels; this is
a compromise between optimizing the spatial resolution ($\sim 385$ pc
aperture$^{-1}$) and the signal-to-noise ratio (S/N), while also mitigating
the effects of the curvature of the spectral axis with respect to the CCD
chip. We calculated the Poisson error on the raw data frames and propagated
the uncertainty through the reduction and analysis.

\subsection{Flux Calibration}
\label{chap3:sec:flux_cal}

Using observations obtained on 29 March 2016 and 25 February 2017, we
performed the relative flux calibration for s3o and s1/s2 using the
spectrophotometric standard stars Hiltner 600 and HR5501,
respectively. No relative flux calibration was performed for s3uv due
to the small wavelength range of interest around the [OII] doublet.

Absolute flux calibration cannot be performed with SALT alone due to the
time-varying collecting area. We calibrated the H$\alpha$ intensities along
the minor axis using calibrated H$\alpha$ narrow-band imaging obtained from
the Hubble Legacy Archive\footnote[3]{\url{https://hla.stsci.edu/}} (PI:
Rossa; Proposal ID: 10416). The observations were taken using ACS/WFC on the
\textit{Hubble Space Telescope} (\textit{HST}) with the f658n filter. We
convolved the image with a two-dimensional Gaussian with full width at half
maximum (FWHM) equal to the average seeing at the SALT site (FWHM = $1.5''$)
using the IDL function \texttt{gauss\char`_smooth}.

To determine the H$\alpha$ calibration factor, we computed the mean
flux over the bandpass for each extracted spectral aperture in
the \textit{HST} data using the PHOTFLAM keyword. We then compared
this to the mean flux in the sky-subtracted SALT spectra convolved
with the \textit{HST} bandpass. We performed the analysis within
$|z| \le 5$ kpc, but we restricted our determination of a median
calibration factor to within $|z| \le 2$ kpc due to declining S/N at
large $z$. The SALT-RSS slit width is sampled by
25 \textit{HST} pixels, a sufficient number to avoid an infinite
aperture correction. We do not account for uncertainty in the flux
calibration in the error budget.

\subsection{Emission-line Detection and Fitting}
\label{chap3:sec:em_line}

We determined the detection limits for emission lines as follows. We
identified statistically significant fluctuations in the spectra by first
fitting and removing any remaining continuum; the fit was performed by masking
emission lines and smoothing the spectrum with a broad Gaussian kernel
($\sigma \gtrsim 15$ \AA). We then divided the spectrum into intervals equal
to the expected FWHM of the emission lines (FWHM = 2.5 - 6 \AA). Within each
interval, we summed the counts per \AA; we then fit the resulting distribution
of integrated intensities with a Gaussian function and determined its standard
deviation. We defined detections to be intervals whose integrated intensities
were statistically significant at the 5$\sigma$ level.

We characterized the intensities, velocities, and velocity dispersions of
detected emission lines by fitting them with single Gaussians using the IDL
function \texttt{gaussfit}. The S/N does not permit reliable identification of
multiple Gaussian components. The least-squares fitting routine also returns
uncertainties on the best-fit parameters. In the Appendix, we show example
spectra from s1, as well as the model fits and their residuals. We also
include measured intensities, velocities, velocity dispersions, and
emission-line ratios in online-only tables.

Due to partial blending of the [OII] doublet in the s3uv observations,
we fit the doublet with a sum of two Gaussians by minimizing the
$\chi^{2}$ statistic over a parameter grid ($\chi^{2}
= \Sigma \frac{(I_{\lambda,obs} -
I_{\lambda,mod})^{2}}{\sigma_{obs}^{2}}$, where $I_{\lambda,obs}$ and
$I_{\lambda,mod}$ are the observed and modeled specific intensities
and $\sigma_{obs}$ is the uncertainty on the observed quantity). We
judge by eye that the blended Gaussian fits are only reliable at
detection levels of at least 7$\sigma$, and thus we only report the
results of these fits. The 1$\sigma$ error bars on the best-fit
parameters were determined from the models within
$\chi^{2} \le \chi^{2}_{min} + \Delta\chi^{2}$, where $\Delta\chi^{2}
= 3.53$ for three free parameters ($\frac{I_{\lambda 3726}}{I_{\lambda
3726,3729}}$, $v_{[OII]}$, and $\sigma_{[OII]}$).

All emission-line widths reported in this paper are the
standard deviations of the Gaussian fits and are corrected for the
instrumental resolution determined from the comparison lamp
spectra. The radial velocities are presented as heliocentric
velocities calculated using the IRAF task
\texttt{astutil.rvcorrect}. Line ratios are corrected for Galactic extinction
using the reddening law of \citet{Cardelli1989} assuming $R_{V} = 3.1$ and
$A(V) = 0.115$ \citep{Schlafly2011}. We do not correct for internal
extinction or for the underlying stellar absorption spectrum.

\section{Results}
\label{chap3:sec:results}

We detect one or more emission lines to spatial extents of $-7.3\ \text{kpc}
\le z \le 8.0\ \text{kpc}$ in s1, to $-8.5\ \text{kpc} \le z \le 10\
\text{kpc}$ in s3o, and to $-11.6\ \text{kpc} \le R \le 13.5\ \text{kpc}$ in
s2. In s3uv, we detect [OII] emission to $-8.5\ \text{kpc} \le z \le 7.3\
\text{kpc}$; detections at the 7$\sigma$ level or higher required for a
reliable fit are restricted to $|z| \le 5.8$ kpc (see \S
\ref{chap3:sec:em_line}). Here, $z$ is the vertical coordinate, where positive
$z$ is on the northeast side of the disk and negative $z$ is on the southwest
side. We choose $z = 0$ to correspond to the main dust lane in the disk, and
we do not correct for inclination angle. $R$, the radial coordinate, is
defined to be positive and negative west and east of the minor axis,
respectively (see labels in Fig. \ref{chap3:fig1}).

\subsection{H$\alpha$ Intensity Profile}
\label{chap3:sec:Ha_int}

We fit the H$\alpha$ intensity, $I_{H\alpha}$, as a function of
height, $z$, along the minor axis. The goal of the fitting is to
determine the eDIG electron density, $n_{e}(z)$, and the exponential
electron scale height, $h_{z,e}$. Since our observations are
integrated in galactocentric radius and $n_{e}(z)$ and $h_{z,e}$ may
depend on $R$, we take the best-fit values of these parameters as
representative of their characteristic values. We perform the fit
along the minor axis because H$\alpha$ imaging indicates a lack of
filamentary structure at this location \citep{Collins2000}, giving us
the best sense of the vertical structure of the diffuse layer.

We exclude observations within $|z| \le 1\ \text{kpc}$ to avoid
contamination by HII regions, dust extinction, and underlying stellar
absorption in the disk. There is evidence for extraplanar dust
in NGC 5775, including PAH and H$_{2}$ emission with scale heights of
$\sim 0.7 - 1.1$ kpc \citep{Rand2011}. Molecular gas and dust are also
detected to $|z| \le 5$ kpc in discrete locations that appear to
coincide with HI supershells \citep{Lee2002, Brar2003}; however, no
such features are seen along the minor axis at the location of
s1. Thus, we assume that the impact of dust extinction on the measured
eDIG scale height and density distribution above $|z| \sim 1$ kpc is
minor.

\begin{figure}[h]
  \epsscale{1.2}\plotone{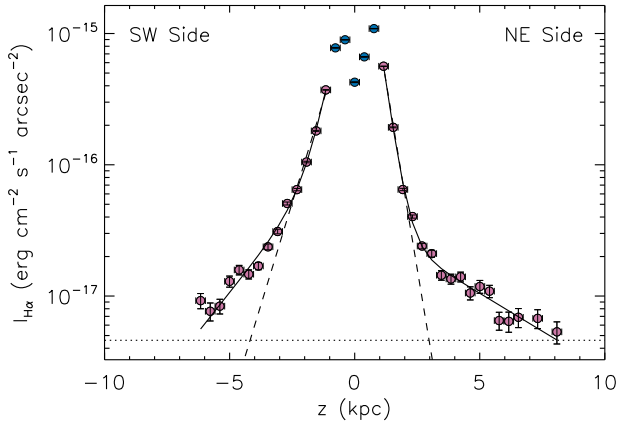}
  \caption{Best-fit, one-component
  (dashed lines) and two-component (solid lines) models of
  $I_{H\alpha}$ as a function of $z$ along the minor axis
  (s1). Best-fit parameters are given in Tab. \ref{chap3:tab2}. Pink
  points ($|z| > 1$ kpc) are included in the fit. The dotted line
  indicates the detection threshold. The slight inclination of
  the galaxy from edge-on may contribute to the asymmetry in the
  $I_{H\alpha}$ profile between the northeast and southwest sides of the
  disk.}
  \label{chap3:fig2}
\end{figure}

We perform the fitting using the IDL function \texttt{mpfitfun}. This routine
performs a Levenberg-Marquardt least-squares fit of a user-supplied function
to a data set. We fit a single exponential function of the form:
\begin{equation}
  I_{H\alpha}(z) = I_{H\alpha}(0)e^{-|z|/h_{z}},
  \label{chap3:eq1}
\end{equation}
as well as a double exponential function given by:
\begin{equation}
  I_{H\alpha}(z) = I_{H\alpha,1}(0)e^{-|z|/h_{z,1}} + I_{H\alpha,2}(0)e^{-|z|/h_{z,2}}.
  \label{chap3:eq2}
\end{equation}
Here, $I_{H\alpha}(0)$ is the intensity of the eDIG layer at $z = 0$ kpc and
$h_{z}$ is the emission scale height. We associate components $1$ and $2$ with
the thick disk and the halo, respectively. As shown in Fig.  \ref{chap3:fig2},
the double exponential provides a much better fit to the intensity profile
than the single exponential at large $z$.

From the $I_{H\alpha}(z)$ profile, we can estimate $n_{e}(z)$ as
follows. $I_{H\alpha}(z)$ depends on $n_{e}(z)$, the electron temperature,
$T_{4}$, and the volume filling factor, $\phi$, according to:
\begin{equation}
  I_{H\alpha}(z) = \frac{\int\phi n_{e}(z)^{2}\text{d}l}{2.75 T_{4}^{0.9}}.
  \label{chap3:eq3}
\end{equation}
Using standard units for the emission measure, $I_{H\alpha}(z)$ is measured in
Rayleighs, $\phi n_{e}^{2}$ is given in $\text{cm}^{-6}$, $T_{4}$ has units of
$10^{4}$ K, and the integral is performed over the line of sight ($\int
\text{d}l = L$ for $L$ in parsecs) \citep[e.g.,][]{Haffner1998}.

If $\phi$ and $n_{e}$ are assumed to be constant along a given line of sight,
then we can express the observable quantity $\phi n_{e}(z)^{2}$ as:
\begin{equation}
  \phi n_{e}(z)^{2} = \frac{I_{H\alpha}(z)(2.75T_{4}^{0.9})}{L}.
  \label{chap3:eq4}
\end{equation}
We take the radial extent of the eDIG layer determined
by \citet{Collins2000}, $R = 14$ kpc, to estimate $L = 2R = 28$ kpc
(note that $R$ has been adjusted for $D$).

We do not include any radial dependence, such as a radial scale length
of the form $n_{e}(R) \propto e^{-R/h_{R}}$, for the derived
densities. Evidence for such a dependence is not clear from existing
imaging and spectroscopy. The H$\alpha$ imaging of \citet{Collins2000}
indicates that the eDIG morphology is largely filamentary, with local
density enhancements dominating over large-scale trends. The radial
density profile derived from Fabry-Perot spectroscopy
by \citet{Heald2006a} also suggests a clumpy rather than smooth
structure. In light of this, we neglect a formal radial density
dependence in our analysis and instead briefly consider the impact of
including an eDIG radial scale length, $h_{R}$, on the success of the
dynamical equilibrium model in \S\ref{chap3:sec:test}.

The best-fit values of $\phi n_{e}(0)^{2}$ and $h_{z,e}$ for the
single, thick disk, and halo components are given in
Tab. \ref{chap3:tab2}. Note that due to the $n_{e}^{2}$ dependence of
$I_{H\alpha}$, the exponential electron scale height is twice the
emission scale height, or $2h_{z} = h_{z,e}$. We quantify the
uncertainties on the best-fit values of $\phi n_{e}(0)^{2}$ and
$h_{z,e}$ by perturbing the values of $I_{H\alpha}$ by increments
randomly drawn from Gaussian distributions with standard deviations
equal to their $1\sigma$ error bars. We then refit the perturbed
intensity profile and calculate the difference between the original
and perturbed best-fit values. We repeat this process until the median
differences converge ($10^{3}$ iterations); these median values are
the uncertainties reported in Tab.
\ref{chap3:tab2}. Note that this approach does not account for systematic
effects such as the deviation of the inclination angle from
$90^{\circ}$ and uncertainty in the flux calibration.

We also compare our best-fit parameters to past measurements of $h_{z,e}$ and
$\phi n_{e}(0)^{2}$ in Tab. \ref{chap3:tab2p2}. Our scale heights are broadly
consistent with the results of these single-component fits, although it is
clear that variation exists as a function of $R$. Our minimum thick-disk scale
height ($h_{z,e} = 0.6$ kpc) and our maximum halo scale height ($h_{z,e} =
7.5$ kpc) bracket existing measurements of the scale height at various
$R$. Thus, by testing a dynamical equilibrium model of each of our thick-disk
and halo components both separately and together, we will assess the success
of this model over a representative range of choices for $h_{z,e}$.

\begin{deluxetable*}{ccccc}[t]
\tabletypesize{\scriptsize}
\tablecolumns{7}
\tablewidth{0pt}
\tablecaption{H$\alpha$ Intensity Profile}
\tablehead{ 
\colhead{Model} &
\colhead{$h_{z,e,1}$} &
\colhead{$\phi n_{e,1}(0)^{2}$} &
\colhead{$h_{z,e,2}$} &
\colhead{$\phi n_{e,2}(0)^{2}$} \\
\colhead{} &
\colhead{(kpc)} &
\colhead{(cm$^{-6}$)} &
\colhead{(kpc)} &
\colhead{(cm$^{-6}$)}
}
\startdata
Thick disk (SW) & 1.4 $\pm$ 0.009 & 0.029 $\pm$ 0.0004 & -- & -- \\
Thick disk (NE) & 0.8 $\pm$ 0.04 & 0.184 $\pm$ 0.003 & -- & -- \\
Thick disk + halo (SW) & 0.8 $\pm$ 0.04 & 0.069 $\pm$ 0.004 & 3.6 $\pm$ 0.2 &
0.003 $\pm$ 0.0004 \\
Thick disk + halo (NE) & 0.6 $\pm$ 0.04 & 0.302 $\pm$ 0.007 & 7.5 $\pm$ 0.4 &
0.0006 $\pm$ 0.00004 
\enddata
\label{chap3:tab2}
\end{deluxetable*}

\begin{deluxetable}{cccc}[]
  \tabletypesize{\scriptsize} \tablecolumns{4} \tablewidth{0pt}
  \tablecaption{Past Measurements of Electron Scale Height and Density}
  \tablehead{ \colhead{$h_{z,e}$} & \colhead{$\phi n_{e}(0)^{2}$} &
    \colhead{Location} &
    \colhead{Reference\tablenotemark{a}} \\
    \colhead{(kpc)} & \colhead{(cm$^{-6}$)} & \colhead{} & \colhead{} }
  \startdata
  1.6 - 1.8 & 0.0065 - 0.01 & $R = 0 - 2$ kpc (-$z$) & [1] \\
  1.8 & 0.0055 & $R = - (8 - 10)$ kpc & [1] \\
  6.0 & - & H$\alpha$ filament (NE side) & [1] \\
  5.0 & - & H$\alpha$ filaments (SE, SW sides) & [1] \\
  4.6 & - & $R = 4.5$ kpc (+$z$) & [2] \\
  4.6 & - & $R = 4.5$ kpc (-$z$) & [2] \\
  5.0 & - & $R = -2.8$ kpc (+$z$) & [2] \\
  3.6 & - & $R = -2.8$ kpc (-$z$) & [2] \\
  1.6 & - & SW side & [3] \\
  1.2 & - & NE side & [3]
\enddata
\tablenotetext{a}{References: [1] - \citealt{Collins2000}; [2] -
  \citealt{Collins2001}; [3] - \citealt{Heald2006a}}
\label{chap3:tab2p2}
\end{deluxetable}

We estimate the total mass in the eDIG layer from the observed density
distribution and an assumed volume filling factor. For $\phi = 1$ ($\phi =
0.1$), we find $M_{eDIG} = 1.1 \times 10^{10}$ \Msun ($M_{eDIG} = 3.5 \times
10^{9}$ \Msun) on the northeast side of the disk and $M_{eDIG} = 8.4 \times
10^{9}$ \Msun ($M_{eDIG} = 2.6 \times 10^{9}$ \Msun) on the southwest
side. $65$\% ($54$\%) of the mass is found in the thick disk on the northeast
(southwest) side. These exceed previous estimates of the masses of eDIG layers
in other galaxies by more than an order of magnitude
\citep[e.g.,][]{Dettmar1990, Boettcher2017}. They are also comparable to the
total HI mass ($M_{HI} = 9.1 \pm 0.6 \times 10^{9}$ \Msun;
\citealt{Irwin1994}) and more than the total hot halo mass ($M_{hot} = 5.3
\times 10^{8}$ \Msun; \citealt{Li2008}).

This suggests that we have overestimated the eDIG mass, implying the need for
a very small volume filling factor. Assuming pressure equilibrium between the
warm and hot phases does indeed imply an eDIG filling factor of $\phi \sim
0.1$ \citep{Li2008}. Additionally, if the eDIG density decreases exponentially
as a function of $R$, then a reasonable choice of radial scale length ($h_{R}
\sim 3 - 5$ kpc) decreases the total mass estimate by $\sim 50$\%.

\subsection{eDIG Properties}
\label{chap3:sec:eDIG_prop}

We use the emission-line ratios to gain a qualitative sense of the physical
conditions in the eDIG layer. The emission-line properties have been
previously studied in detail by \citet{Tullmann2000}, \citet{Collins2001},
\citet{Otte2002}, and others; we confirm their findings at the location of s3,
and expand on their results at the locations of s1 and s2. Note that internal
extinction and underlying stellar absorption may affect observed line ratios
at small $z$; however, we emphasize interpretation of the line ratios at $|z|
\ge 1$ kpc, where we assume these effects are minimal.

The [OII]$\lambda$3727/H$\alpha$, [NII]$\lambda$6583/H$\alpha$, and
[SII]$\lambda$6717/H$\alpha$ line ratios depend on temperature,
ionization fraction, and elemental abundance as
follows \citep{Haffner1999, Osterbrock2006}:
\begin{equation}
  \frac{I([OII]\lambda 3727)}{I(H\alpha)} = (4.31 \times
  10^{5})T_{4}^{0.4}e^{-3.87/T_{4}} \bigg(\frac{O^{+}}{O}\bigg)
  \bigg(\frac{O}{H}\bigg) \bigg(\frac{H^{+}}{H}\bigg)^{-1},
  \label{chap3:eq5}
\end{equation}
\begin{equation}
  \frac{I([NII]\lambda 6583)}{I(H\alpha)} = (1.62 \times
  10^{5})T_{4}^{0.4}e^{-2.18/T_{4}} \bigg(\frac{N^{+}}{N}\bigg)
  \bigg(\frac{N}{H}\bigg) \bigg(\frac{H^{+}}{H}\bigg)^{-1},
  \label{chap3:eq6}
\end{equation}
and
\begin{equation}
  \frac{I([SII]\lambda 6717)}{I(H\alpha)} = (7.67 \times
  10^{5})T_{4}^{0.307}e^{-2.14/T_{4}} \bigg(\frac{S^{+}}{S}\bigg)
  \bigg(\frac{S}{H}\bigg) \bigg(\frac{H^{+}}{H}\bigg)^{-1}.
  \label{chap3:eq7}
\end{equation}
We assume $H^{+}/H \sim 1$ and $N^{+}/N \sim 0.8$ under typical DIG conditions
\citep{Reynolds1998, Sembach2000}. We also assume $O^{+}/O = N^{+}/N$ given
their similar first ionization potentials. Though the relatively high
second ionization potentials of $N$ and $O$ (29.6 and 35.1 eV,
respectively) suggest that $N^{++}/N$ and $O^{++}/O$ are low in the
DIG, the lower second ionization potential of $S$ (23.3 eV) means that
$S^{++}/S$ may be considerable. Thus, we consider models with a range
of $S^{+}/S$ here. We adopt the solar photospheric abundances
of \citet{Asplund2009}: $N/H = 6.8 \times 10^{-5}$, $S/H =
1.3 \times 10^{-5}$, and $O/H = 4.9 \times 10^{-4}$. We assume that
the emission from different atomic species arises co-spatially, and
that the temperature, ionization, and abundances do not vary
significantly along a given line of sight.

\begin{figure}[h]
  \epsscale{1.2}\plotone{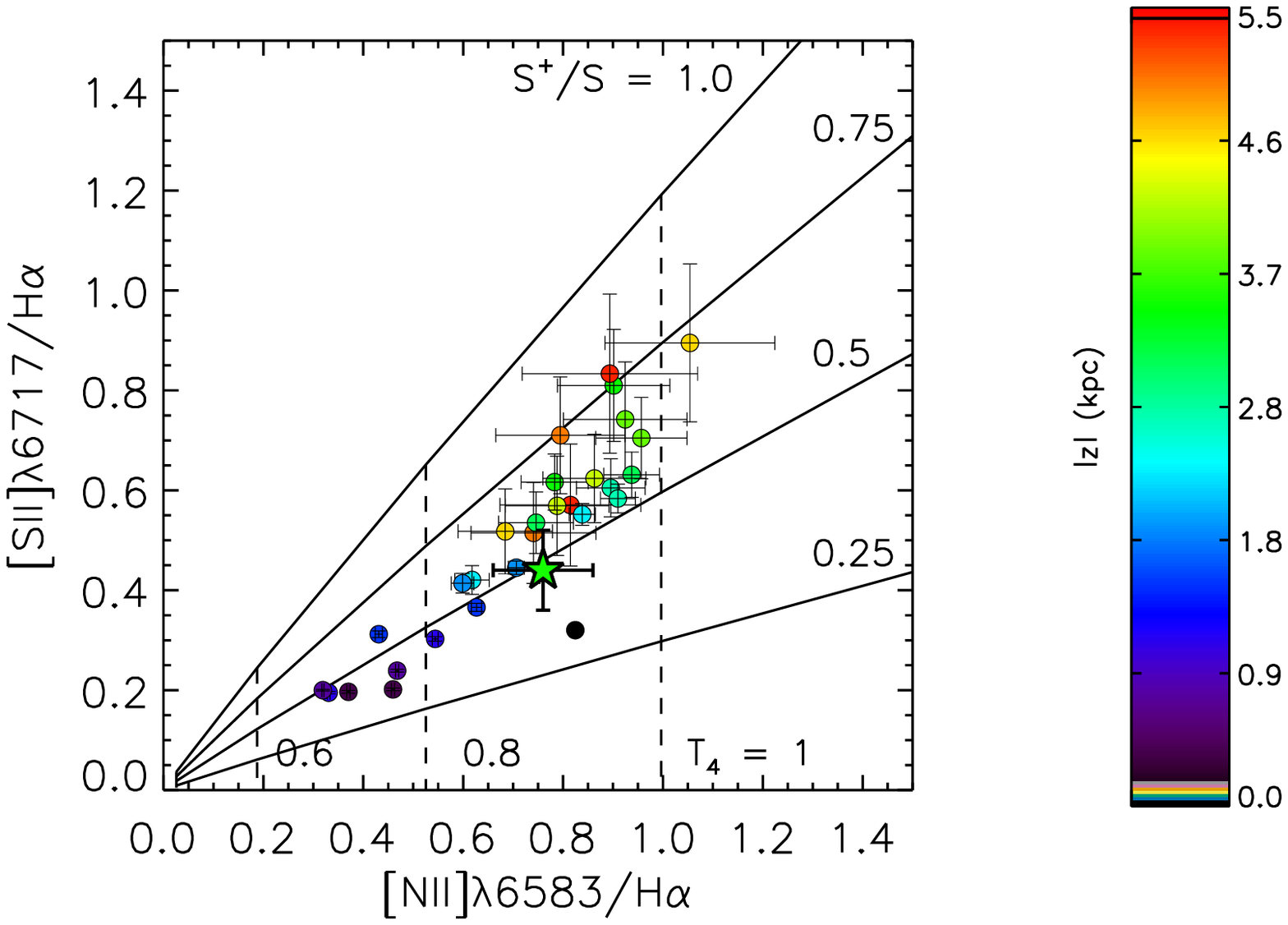}
  \epsscale{1.2}\plotone{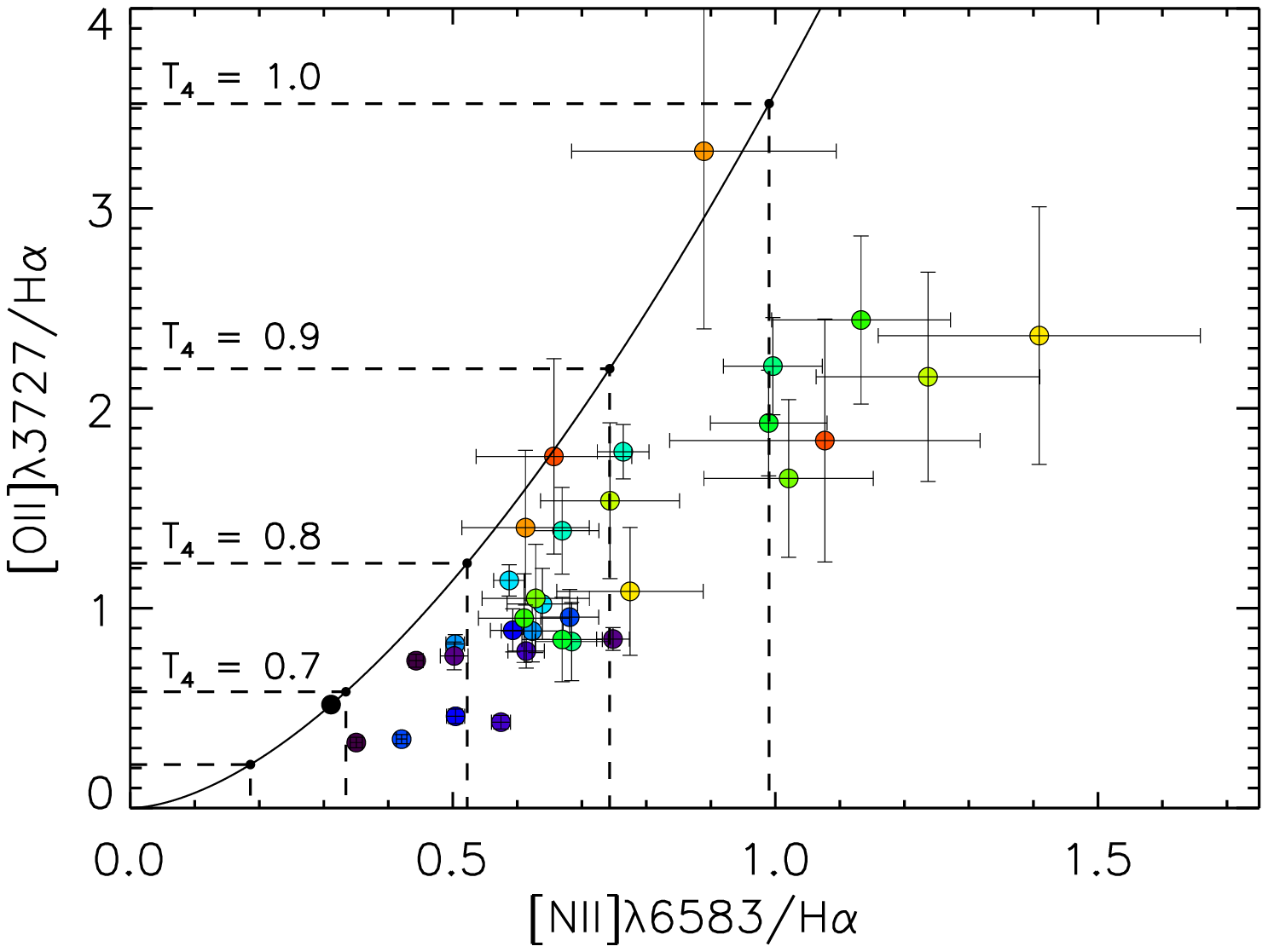}
  \caption{Top: [NII]$\lambda$6583/H$\alpha$ vs. [SII]$\lambda$6717/H$\alpha$
  color-coded by $|z|$ along the minor axis (s1). The median and
  median absolute deviation of the line ratios from s2 are shown by
  the star and accompanying error bar. The solid lines indicate the
  expected line ratios for a range of $T_{4}$ at fixed $S^{+}/S$. The
  dashed lines mark locations of constant $T_{4}$. The rising line
  ratios suggest an increase in $T_{4}$ with $|z|$.  Bottom:
  [NII]$\lambda$6583/H$\alpha$ vs. [OII]$\lambda$3727/H$\alpha$
  color-coded by $|z|$ along the filament studied in s3. The observed
  line ratios generally fall below the predicted ratios at
  all $T_{4}$, likely due to selective internal extinction and/or
  abundance effects. As in the top panel, these results suggest
  increasing $T_{4}$ with $|z|$ and an eDIG temperature of $0.8 \leq
  T_{4} \leq 1.0$.}  \label{chap3:fig3}
\end{figure}

Given these assumptions, we compare the observed and expected emission-line
ratios for a range of $T_{4}$ and $S^{+}/S$. In the top panel of Fig.
\ref{chap3:fig3}, we plot [NII]$\lambda$6583/H$\alpha$
vs. [SII]$\lambda$6717/H$\alpha$ observed along the minor axis in
s1. We also show the median and median absolute deviation of these
line ratios observed in s2; these ratios do not show a strong trend
with slit position at this location. Both line ratios clearly increase
as a function of $z$, a trend commonly observed in the eDIG of edge-on
galaxies and often interpreted as evidence of increasing $T_{4}$ with
$z$ \citep[e.g.,][]{Haffner1999}. At $|z| > 2$ kpc, the line ratios
suggest a temperature of $T_{4} = 0.8 - 1.0$ and an ionization
fraction of $S^{+}/S \sim 0.5 - 0.75$. There is also a trend
toward increasing $S^{+}/S$ with $z$ as $S^{++}$ becomes $S^{+}$ in a
more dilute radiation field.

In the bottom panel of Fig. \ref{chap3:fig3}, we plot
[NII]$\lambda$6583/H$\alpha$ vs. [OII]$\lambda$3727/H$\alpha$ observed
along the filament in s3. Since the [OII] doublet is not resolved in
these observations, the [OII] intensity is integrated over both
lines. As $N^{+}/N$ and $O^{+}/O$ are thought to be comparable in the
DIG, these line ratios depend only on temperature for fixed
abundances. We compare the observed and predicted line ratios for a
range of temperatures, and we find that the observed ratios fall at
lower values of [OII]$\lambda$3727/H$\alpha$ than expected for a given
value of [NII]$\lambda$6583/H$\alpha$. This may be due to selective
internal extinction reddening the [OII]$\lambda$3727/H$\alpha$ line
ratios within the dusty, thick-disk ISM. Additionally, both line
ratios may be affected by underlying stellar absorption that reduces
the measured H$\alpha$ intensity.  However, it is not clear that the
observed and expected line ratios converge at large $z$ where these
effects are likely minimal. Thus, it is possible that enhancement in
the $N$ abundance with respect to $O$ is also responsible.

\begin{figure}[h]
  \epsscale{1.2}\plotone{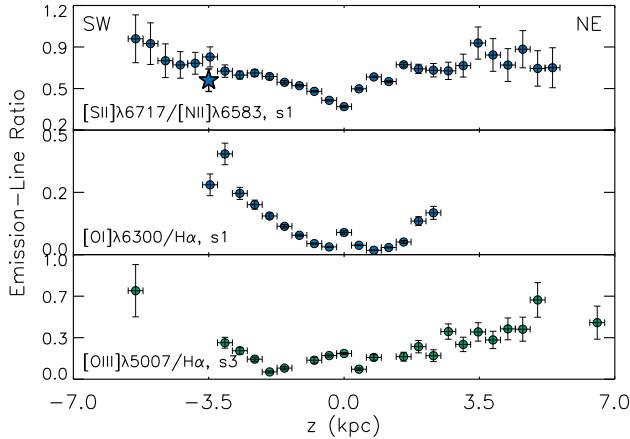}
  \caption{[SII]$\lambda$6717/[NII]$\lambda$6583 (top; s1),
    [OI]$\lambda$6300/H$\alpha$ (middle; s1), and
    [OIII]$\lambda$5007/H$\alpha$ (bottom; s3) as functions of $z$. The
    observed increase in [OI]/H$\alpha$ and [OIII]/H$\alpha$ suggests a
    supplemental heating and/or ionization source such as shocks or turbulent
    mixing layers \citep{Collins2001}.}
  \label{chap3:fig4}
\end{figure}

From Fig. \ref{chap3:fig3}, we adopt an eDIG temperature of $T_{4} = 0.9$ for
the rest of this paper. As we will see in \S\ref{chap3:sec:eDIG_vel_disp}, the
turbulent velocity dispersion far exceeds the thermal velocity dispersion, and
thus uncertainty in the eDIG temperature on the order of $10 - 20\%$ is
unimportant for our dynamical equilibrium study. Out of the disk, the observed
[SII]$\lambda$6717/[SII]$\lambda$6731 $\sim 1.5$ in s1 and
[OII]$\lambda$3729/[OII]$\lambda$3726 $\sim 1.5$ in s3uv demonstrate that the
eDIG is in the low-density limit as expected \citep[e.g.,][]{Osterbrock2006}.

The emission-line ratios of eDIG layers are also of interest due to the
challenge that they present to photoionization models
\citep[e.g.,][]{Collins2001}. We comment briefly on that here. In Fig.
\ref{chap3:fig4}, we show [SII]$\lambda$6717/[NII]$\lambda$6583 and
[OI]$\lambda$6300/H$\alpha$ as functions of $z$ along the minor axis (s1) and
[OIII]$\lambda$5007/H$\alpha$ as a function of $z$ in s3. (Note that detection
of the [OI] line in s3 is impeded by sky-line residuals. We also choose to
take the ratio of [OIII] to H$\alpha$ instead of the more traditional H$\beta$
because the latter is not detected at large $z$ and is contaminated by the
underlying stellar absorption line at low $z$.)

In a photoionization model, we expect
[SII]$\lambda$6717/[NII]$\lambda$6583 to increase with distance from
the ionizing source as $S^{++} \rightarrow S^{+}$. We do observe an
increase in [SII]/[NII] as a function of $z$. However, the rise in
[OIII]$\lambda$5007/H$\alpha$ with $z$ is associated with a need for
an additional source of heating and/or ionization, such as shocks or
turbulent mixing layers
\citep[e.g.,][]{Collins2001}. The presence of shocks is further suggested by
the rising [OI]$\lambda$6300/H$\alpha$. Since the neutral fractions of O and H
are coupled via charge exchange, the presence of [OI] emission suggests a
multiphase medium, as is found in shock-compressed regions
\citep[e.g.,][]{Collins2001}. The need for an additional source of heating
and/or ionization that behaves differently than photoionization will be
relevant as we consider the dynamical state of the eDIG layer in Sections
\S\ref{chap3:sec:dyn_eq} and \S\ref{chap3:sec:disc}.

From Fig. \ref{chap3:fig4}, we can also verify the validity of our
assumption that $N^{+}/N$ is high by confirming that $N^{++}/N$ must
be low. Due to the similarly high second ionization potentials of $O$
and $N$, a low $O^{++}/O$ suggests a similarly low $N^{++}/N$. Our
observed [OIII]$\lambda$5007/H$\alpha$ line ratios and the electron
temperature deduced from Fig. \ref{chap3:fig3} ($T_{4} = 0.9$) imply a
median value of $O^{++}/O = 0.1$, confirming a low value of
$N^{++}/N$. (We thank R. Rand for this suggestion.)

\subsection{eDIG Kinematics}
\label{chap3:sec:eDIG_kin}

\subsubsection{Radial Velocities}
\label{chap3:sec:eDIG_vel}

\begin{figure}[h]
  \epsscale{1.2}\plotone{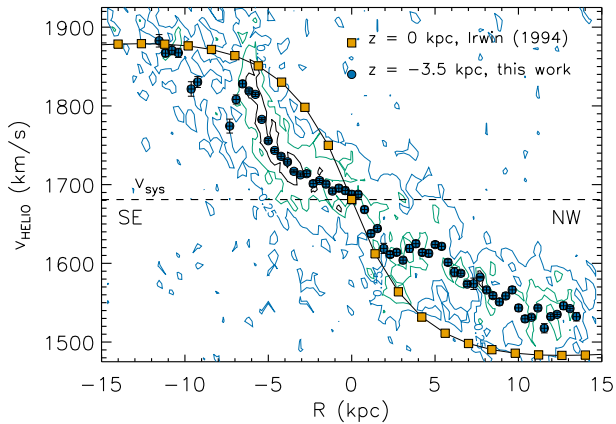}
  \caption{Line-of-sight, heliocentric H$\alpha$ velocities as a function of
    $R$ observed in s2 ($z = -3.5$ kpc; blue) compared to a model HI rotation
    curve from \citet{Irwin1994} ($z = 0$ kpc; yellow). The blue, green, and
    black contours illustrate the position-velocity diagram at 25\%, 50\%, and
    75\% of the maximum H$\alpha$ intensity, respectively. The tendency of the
    eDIG layer towards systemic velocity is characteristic of lagging gaseous
    halos; we find median velocity offsets of $\Delta v = -20.3\ \text{\kms}\
    \text{kpc}^{-1}$ and $\Delta v = -15.4\ \text{\kms}\ \text{kpc}^{-1}$ west
    and east of the minor axis, respectively.}
  \label{chap3:fig5}
\end{figure}

The heliocentric, line-of-sight H$\alpha$ velocity, $v_{H\alpha}$, observed in
s2 is shown in Fig. \ref{chap3:fig5}. The H$\alpha$ rotation curve at $z =
-3.5$ kpc is compared to a model of the HI rotation curve at $z = 0$ kpc from
\citet{Irwin1994}. West and east of the minor axis, $v_{H\alpha}$ is closer to
the systemic velocity than $v_{HI}$ by a median $\Delta v = 71$ \kms and
$\Delta v = 54$ \kms at $|R| \ge 3$ kpc, respectively. The eDIG layer thus
shows the signature of a lagging gaseous halo that arises as a consequence of
the conservation of angular momentum in a disk-halo flow. The larger
asymmetric drift of the warmer gas may also contribute to the velocity
discrepancy between the two phases.

If the line-of-sight velocity samples gas at the tangent point and is thus a
measure of the rotational velocity, then this implies a rotational velocity
gradient of $\Delta v = -20.3\ \text{\kms}\ \text{kpc}^{-1}$ and $\Delta v =
-15.4\ \text{\kms}\ \text{kpc}^{-1}$ for positive and negative $R$,
respectively. However, this is significantly steeper than the $-1\
\text{\kms}\ \text{arcsec}^{-1}$ ($-7\ \text{\kms}\ \text{kpc}^{-1}$)
determined by \citet{Heald2006a} for this galaxy. These authors demonstrate
that quantification of the true rotational velocity gradient is dependent on
robust modeling of a three-dimensional, rotating disk in which the density
distribution and rotation curve are allowed to vary with distance from the
midplane. Thus, we avoid over-interpreting the relatively steep velocity
gradients observed here, as they may be biased by local filamentary structure
and may not reflect the true rotational velocity of the eDIG layer.

\begin{figure}[h]
  \epsscale{1.2}\plotone{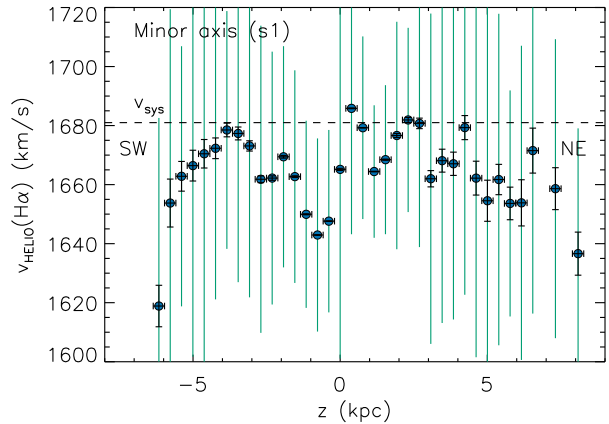}
  \epsscale{1.2}\plotone{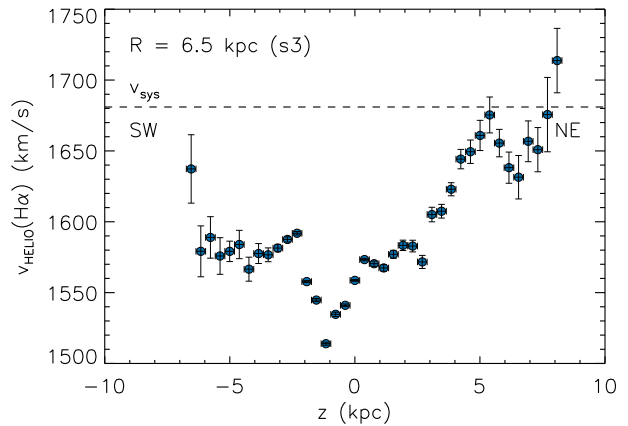}
  \caption{Line-of-sight, heliocentric H$\alpha$ velocities as a function of
    $z$ observed on the minor axis (top; s1) and at $R = 6.5$ kpc (bottom;
    s3). At top, the green bars indicate the emission-line widths,
    $\sigma_{H\alpha}$ (see \S\ref{chap3:sec:eDIG_vel_disp}). At bottom, the
    observed velocity gradients of $\Delta v \sim -25\ \text{\kms}\
    \text{kpc}^{-1}$ (northeast) and $\Delta v \sim -6\ \text{\kms}\
    \text{kpc}^{-1}$ (southwest) are consistent with a previously-observed
    lagging halo \citep[e.g.,][]{Rand2000, Heald2006a}, though we do not
    attempt to quantify the rotational velocity gradient here.}
  \label{chap3:fig6}
\end{figure}

The top panel of Fig. \ref{chap3:fig6} shows $v_{H\alpha}$ as a function of
$z$ observed on the minor axis in s1. There is a spread in $v_{H\alpha}$ at
small $z$ before it approaches $v_{sys}$ at larger $z$. At the largest $z$,
$v_{H\alpha}$ is blueshifted with respect to $v_{sys}$, moving away from the
systemic velocity at increasing distances from the midplane. If the eDIG layer
is cylindrically rotating, then we would expect $v_{H\alpha} \sim v_{sys}$
along the minor axis. Although the spread in $v_{H\alpha}$ at small $z$ is
likely due to the deviation of the inclination angle from $i = 90^{\circ}$,
the blueshifted velocities observed at large $z$ merit further discussion in
\S\ref{chap3:sec:disc}.

The bottom panel of Fig. \ref{chap3:fig6} again shows $v_{H\alpha}$ as a
function of $z$, here observed perpendicular to the disk at $R = 6.5$ kpc in
s3. As previously found at this location by \citet{Rand2000}, $v_{H\alpha}$
climbs steadily from the rotational velocity in the disk, $v_{disk} \sim 1550$
\kms, to the systemic velocity, $v_{sys} = 1681$ \kms, as $z$ increases
($v_{sys}$ is derived using the optical convention from the HI redshift of
\citealt{deVaucouleurs1991}). $v_{H\alpha}$ reaches $v_{sys}$ around $z = 5$
kpc, displaying small variations around the systemic velocity at larger $z$. A
similar trend is seen on the southwest side of the galaxy, although
$v_{H\alpha}$ approaches $v_{sys}$ at a slower rate and does not reach
$v_{sys}$ over the spatial extent of our emission-line detections. Within $|z|
\le 5$ kpc, the velocity gradients are approximately $\Delta v \sim -25\
\text{\kms}\ \text{kpc}^{-1}$ and $\Delta v \sim -6\ \text{\kms}\
\text{kpc}^{-1}$ on the northeast and southwest sides, respectively. While
this is consistent with the presence of a rotational velocity gradient, we
again caution against quantifying this gradient without additional modeling.

It is interesting to ask whether there is any evidence of perturbations to the
line-of-sight velocities due to the interaction with NGC 5774. The most likely
place to find such evidence would be in s2 where the slit extends across the
HI bridge between the galaxies. However, we only detect emission to $R = 13.5$
kpc on this side of the galaxy, suggesting that any ionized gas in the bridge
has too low surface brightness to be detected here. The blueshifted velocities
at large $z$ on the minor axis are the subject of further discussion in
\S\ref{chap3:sec:disc}, including their possible relationship to an
interaction.

\subsubsection{Turbulent Velocity Dispersion}
\label{chap3:sec:eDIG_vel_disp}

\begin{figure}[h]
  \epsscale{1.2}\plotone{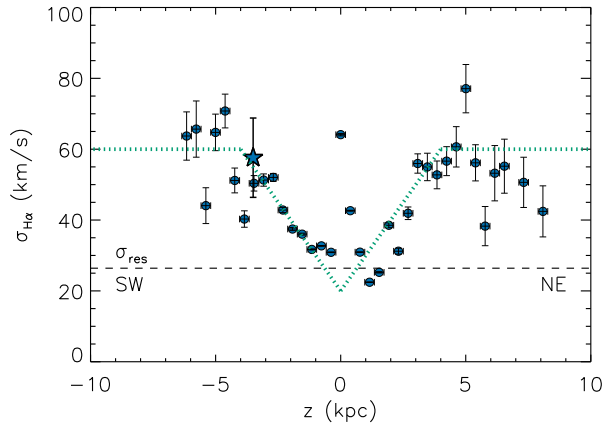}
  \caption{H$\alpha$ line widths, $\sigma_{H\alpha}$, as a function of $z$ on
    the minor axis (s1), where contributions from rotational velocity are
    minimized. $\sigma_{H\alpha}$ is corrected for the spectral resolution,
    indicated by the dashed black line. The median and median absolute
    deviation from s2 are shown by the star and accompanying error
    bar. $\sigma_{H\alpha}$ displays an increase with $z$ within several
    kiloparsecs of the disk, the first clear detection of such a trend in an
    eDIG layer. The dotted green line shows the functional form of $\sigma(z)$
    used in \S\ref{chap3:sec:dyn_eq}.}
  \label{chap3:fig7}
\end{figure}

The H$\alpha$ emission-line widths, $\sigma_{H\alpha}$, are shown as a
function of $z$ along the minor axis in Fig. \ref{chap3:fig7}. These line
widths refer to the standard deviation of the Gaussian velocity
distribution. In the disk, where we expect $\sigma_{H\alpha}$ to be
characteristic of HII regions ($\sigma_{HII} \sim 20$ km/s), the line width is
limited by the spectral resolution ($\sigma_{res} \sim 26$ \kms). However,
$\sigma_{H\alpha}$ smoothly increases with $z$, reaching values as high as
$\sim 70$ \kms by $|z| = 5$ kpc. The median and median absolute deviation
observed in s2 ($z = -3.5$ kpc) is $\sigma_{H\alpha} = 58 \pm 11$ \kms;
there is no obvious trend in $\sigma_{H\alpha}$ with $R$.

This is the first clear evidence of an increase in emission-line
widths as a function of height above the disk in an eDIG layer. The
line widths appear to rise within $|z| \le 4$ kpc and then plateau at
some characteristic value. As discussed in \S\ref{chap3:sec:Ha_int},
the impact of dust extinction on the observed velocity dispersion is
likely to be small, as the increase in $\sigma_{H\alpha}$ with $z$ is
observed at several dust scale heights above the disk. Several factors
contribute to the line width, including thermal, turbulent, and
noncircular motions (note that we have already corrected for the
spectral resolution, and rotational motions are minimized on the minor
axis). The thermal line width of a gas with $T_{4} \sim 0.9$ is small
compared to the observed line widths ($\sigma_{th} \sim 10$ \kms); in
the absence of information about contributions from noncircular
motions, we take the observed line widths as an upper limit on the
turbulent velocity dispersion in the eDIG layer. Here, ``turbulence''
refers to any random gas motions, including the cloud-cloud velocity
dispersion. We discuss possible origins for the $z$-dependence of
$\sigma_{H\alpha}$ in \S\ref{chap3:sec:disc}.

\section{A Dynamical Equilibrium Model}
\label{chap3:sec:dyn_eq}

We now ask whether the eDIG layer in NGC 5775 is well-represented by a
dynamical equilibrium model - that is, whether the thermal, turbulent,
magnetic field, and cosmic-ray pressure gradients are sufficient to satisfy
the equation of hydrostatic equilibrium. (We favor the term ``dynamical
equilibrium'' over ``hydrostatic equilibrium'' due to the inclusion of
turbulent motions in the relevant sources of pressure support.) We follow the
approach described in \citet{Boettcher2016}, which we summarize here.

We solve the hydrostatic equilibrium equation for an eDIG layer in pressure
balance given by:
\begin{equation}
  \frac{dP(z,R)}{dz} = -\frac{d\Phi(z,R)}{dz}\rho(z,R).
  \label{chap3:eq8}
\end{equation}
Here, $P(z,R)$ and $\rho(z,R)$ are the gas pressure and density, respectively,
and $\frac{d\Phi(z,R)}{dz} = g_{z}(z,R)$ is the gravitational acceleration
derived from the galactic gravitational potential, $\Phi(z,R)$. We discuss the
mass model used to determine the galactic gravitational potential in the
Appendix.

We consider contributions from gas pressure, $P_{g}$, magnetic field pressure,
$P_{B}$, and cosmic-ray pressure, $P_{cr}$ (note that we do not consider
magnetic tension):
\begin{equation}
  P(z,R) = P_{g} + P_{B} + P_{cr}.
  \label{chap3:eq9}
\end{equation}
The gas pressure is dependent on the velocity dispersion, $\sigma$, that in
turn depends on both thermal ($\sigma_{th}$) and turbulent ($\sigma_{turb}$)
motions:
\begin{equation}
  P_{g}(z,R) = \sigma(z)^{2}\rho(z,R), \sigma(z)^{2} = \sigma_{th}^{2} +
  \sigma_{turb}^{2}.
  \label{chap3:eq10}
\end{equation}

Our goal is to determine whether we can satisfy Eq. (\ref{chap3:eq8}) given
the observed gas density distribution, $\rho(z)$, and the observed pressure
gradients, $\frac{dP}{dz} = \frac{dP_{g}}{dz} + \frac{dP_{B}}{dz} +
\frac{dP_{cr}}{dz}$. In \S\ref{chap3:sec:therm} - \S\ref{chap3:sec:cr} below,
we derive expressions for the relevant pressure gradients, expressed
for a single component of the density distribution. As the pressure
gradients are derived from observations integrated across the disk, we
assume that only the gravitational acceleration has radial
dependence. Thus, we take these gradients as representative of the
average vertical pressure support in the eDIG layer. The lack
of significant variation in the observed emission-line widths as a
function of projected galactocentric radius in s2 supports this
approach; note, however, that the line widths can be broadened by
integration through a rotating medium away from the minor axis.

\subsection{Thermal Pressure Gradient}
\label{chap3:sec:therm}

From Eq. (\ref{chap3:eq10}), the thermal gas pressure is given by $P_{th}(z) =
\sigma_{th}^{2}\rho(z)$, where the thermal velocity dispersion can be
expressed as $\sigma_{th} = \sqrt{\frac{kT}{\alpha m_{p}}}$. Here, $k$ is the
Boltzmann constant, $m_{p}$ is the proton mass, and $\alpha$ is a scaling
factor. We take $\alpha = 0.62$ assuming that the gas is 9\% He by number,
with 100\% and 70\% of the H and He ionized, respectively \citep{Rand1997,
  Rand1998}. For a gas with $T_{4} = 0.9$ (\S\ref{chap3:sec:eDIG_prop}), we
find $\sigma_{th} = 11$ \kms. Assuming that $\sigma_{th}$ does not change
substantially with $z$, then the thermal pressure gradient is given by:
\begin{equation}
  \frac{dP_{th}}{dz} = -\frac{\sigma_{th}^{2}}{h_{z}}\rho(z).
  \label{chap3:eq11}
\end{equation}

\subsection{Turbulent Pressure Gradient}
\label{chap3:sec:turb}

Like the thermal gas pressure, the turbulent gas pressure is given by
$P_{turb}(z) = \sigma_{turb}(z)^{2}\rho(z)$, except here we allow the velocity
dispersion to vary with $z$. By eye, we fit the following functional form to
$\sigma_{turb}(z)$, assuming isotropic turbulence (see the green dotted line
in Fig. \ref{chap3:fig7}):
\begin{equation}
  \sigma_{turb}(z) =
  \begin{cases}
    \sigma_{0} + m_{\sigma}z, & \text{if}\ 0\ \text{kpc} \le |z| \le 4\ \text{kpc} \\
    \sigma_{1}, & \text{if}\ |z| > 4\ \text{kpc}.
  \end{cases}
  \label{chap3:eq12}
\end{equation}
Here, we use $\sigma_{0} = 20$ \kms, $m_{\sigma}$ = 10 \kms kpc$^{-1}$, and
$\sigma_{1} = 60$ \kms.

Note again that the observed $\sigma(z)$ is an upper limit to the true
$\sigma_{turb}(z)$ due to possible contributions from noncircular motions. We
have also not corrected the observations for the (small) contribution from
thermal motions. This yields the following turbulent pressure gradient:
\begin{equation}
  \frac{dP_{turb}}{dz} =
  \begin{cases}
    2m_{\sigma}\sigma_{turb}(z)\rho(z) - \frac{\sigma_{turb}(z)^{2}}{h_{z}}\rho(z), & \text{if}\ 0\ \text{kpc} \le |z| \le 4\ \text{kpc} \\
    -\frac{\sigma_{1}^{2}}{h_{z}}\rho(z), & \text{if}\ |z| > 4\ \text{kpc}.
  \end{cases}
  \label{chap3:eq13}
\end{equation}

\subsection{Magnetic Pressure Gradient}
\label{chap3:sec:mag}

To estimate the magnetic field and cosmic-ray pressure gradients, we turn to
the radio continuum observations discussed in
\S\ref{chap3:sec:intro}. \citet{Krause2018} characterize the magnetic field
strengths in the disks and the synchrotron scale heights of galaxies in the
CHANG-ES sample, including thin disk and halo components. For NGC 5775, they
find a disk field strength of $B_{0} = 14.8\ \mu$G and halo synchrotron scale
heights of $h_{z,syn} = 1.46\ \pm\ 0.13$ kpc (C-band D-array; centered at 6
GHz) and $h_{z,syn} = 1.98\ \pm\ 0.35$ kpc (L-band C-array; centered at 1.5
GHz). We discuss the thin synchrotron disk in \S\ref{chap3:sec:disc}.

To estimate the magnetic pressure gradient in the halo of NGC 5775, we assume
energy equipartition between the magnetic field and cosmic-ray energy
densities (see \S\ref{chap3:sec:disc} for a discussion of the merits of this
assumption). Under this assumption, the synchrotron scale height is related to
the magnetic field scale height via the nonthermal spectral index: $h_{z,B}
\sim h_{z,syn}(3 + \alpha)$ \citep{Beck2015}. Note that $h_{z,B}$ may be even
larger due to energy losses of cosmic-ray electrons lowering the synchrotron
scale height.

For $\alpha = 1.09$ \citep{Krause2018}, $h_{z,B} = 6$ kpc and $h_{z,B} = 8.1$
kpc determined from the 6 GHz and 1.5 GHz observations, respectively. We adopt
the latter value here, as lower frequency emission arises from lower energy
cosmic rays that undergo fewer losses than their higher energy
counterparts. The vertical distribution of lower energy cosmic-ray electrons
therefore better approximates the distribution of (lossless) cosmic-ray
protons; the cosmic-ray pressure, of interest in \S\ref{chap3:sec:cr}, arises
from this latter component.

We assume a simple, plane-parallel geometry for the halo component of the
magnetic field, where:
\begin{equation}
  B(z) = B_{0}e^{-|z|/h_{z,B}}.
  \label{chap3:eq14}
\end{equation}
Therefore, given $P_{B} = \frac{B(z)^{2}}{8\pi}$, the magnetic pressure gradient can be written as:
\begin{equation}
  \frac{dP_{B}(z)}{dz} = -\frac{2}{h_{B}}P_{B}.
  \label{chap3:eq15}
\end{equation}
The observed ``X-shaped'' morphology of the halo magnetic field suggests that
it becomes increasingly vertical at large $z$ \citep{Soida2011}. However, the
likely presence of a significant turbulent component to the field mitigates
the effect of the field geometry on the magnetic pressure gradient.
 
\subsection{Cosmic-ray Pressure Gradient}
\label{chap3:sec:cr}

To determine the cosmic-ray pressure gradient, we again assume equipartition
between the magnetic field and cosmic-ray energy densities. Since the majority
of the cosmic-ray energy density is believed to be due to only mildly
relativistic protons \citep{Ferriere2001}, we take $P_{cr} = 0.45U_{cr} =
0.45U_{B}$ and find:
\begin{equation}
  \frac{dP_{cr}(z)}{dz} = 0.45\frac{dP_{B}(z)}{dz}.
  \label{chap3:eq16}
\end{equation}

\subsection{Testing Dynamical Equilibrium}
\label{chap3:sec:test}

We first test the dynamical equilibrium model for an initial choice of
parameters: a volume filling factor of $\phi = 1$ (no gas clumping) and the
magnetic field properties discussed in \S\ref{chap3:sec:mag}. We restrict our
analysis to $|z| \ge 1$ kpc, as the nature of the gas density profile within
$|z| < 1$ kpc is not well characterized due to obscuration by dust and HII
regions and the slight inclination of the disk from edge-on. We thus assume
that some boundary condition governs the meeting of the eDIG layer with the
thin-disk ISM at small $z$. We analyze each side of the galaxy individually,
assessing the dynamical state of the thick disk and halo both together and
separately.

We calculate the thermal, turbulent, magnetic field, and cosmic-ray pressure
gradients and compare the total observed gradient with the required gradient
(in other words, with the right-hand side of Eq. \ref{chap3:eq8}). We consider
the model satisfied if the observed pressure support equals or exceeds the
required support at $|z| \ge 1$ kpc. We do so at every galactocentric radius
between the center and the edge of the eDIG layer ($0\ \text{kpc} \le R \le
14\ \text{kpc}$, in steps of $\Delta R = 1$ kpc), and we determine the minimum
radius at which the dynamical equilibrium model is satisfied, $R_{eq}$.

\begin{figure*}[h]
  \epsscale{1.1}\plottwo{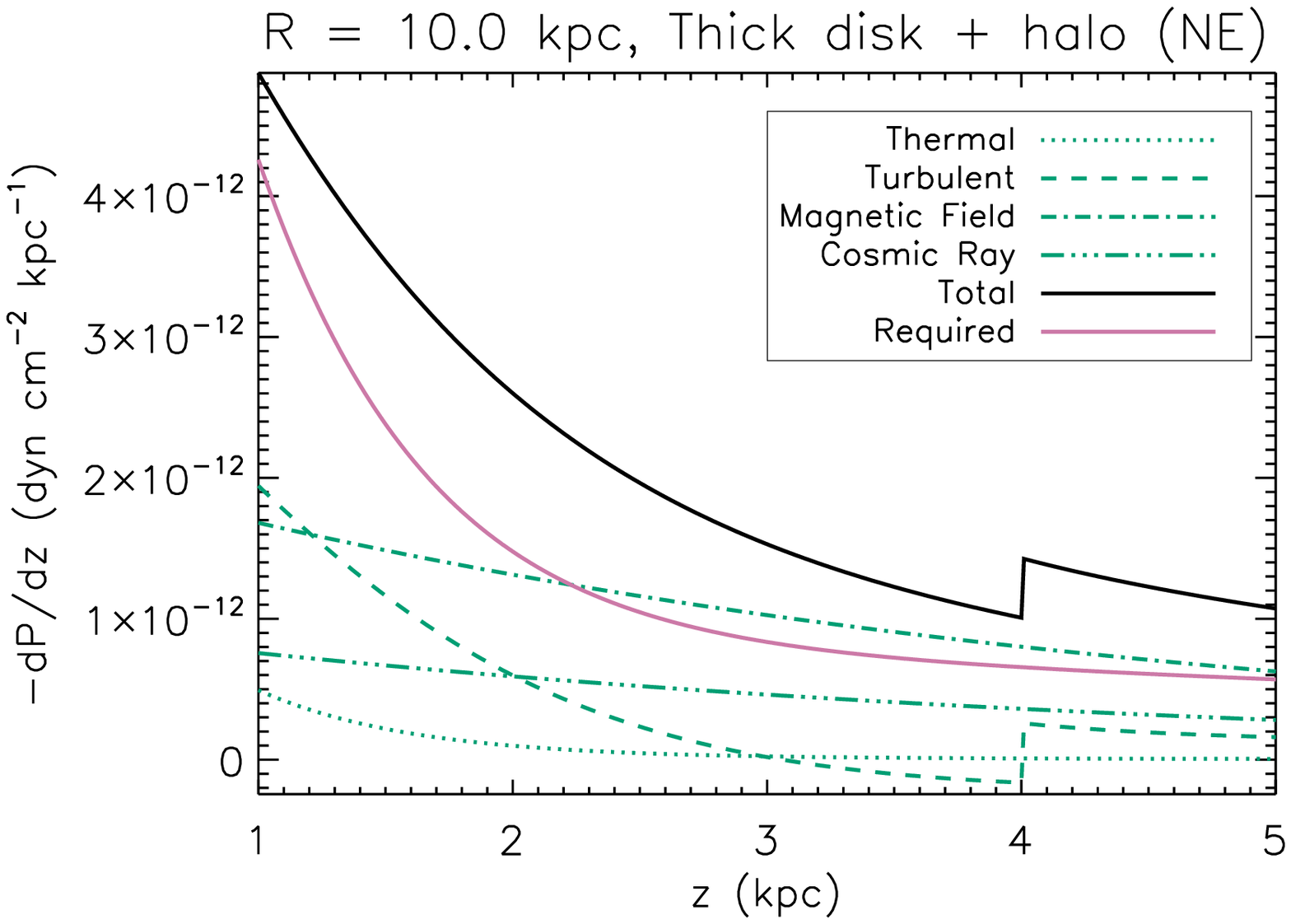}{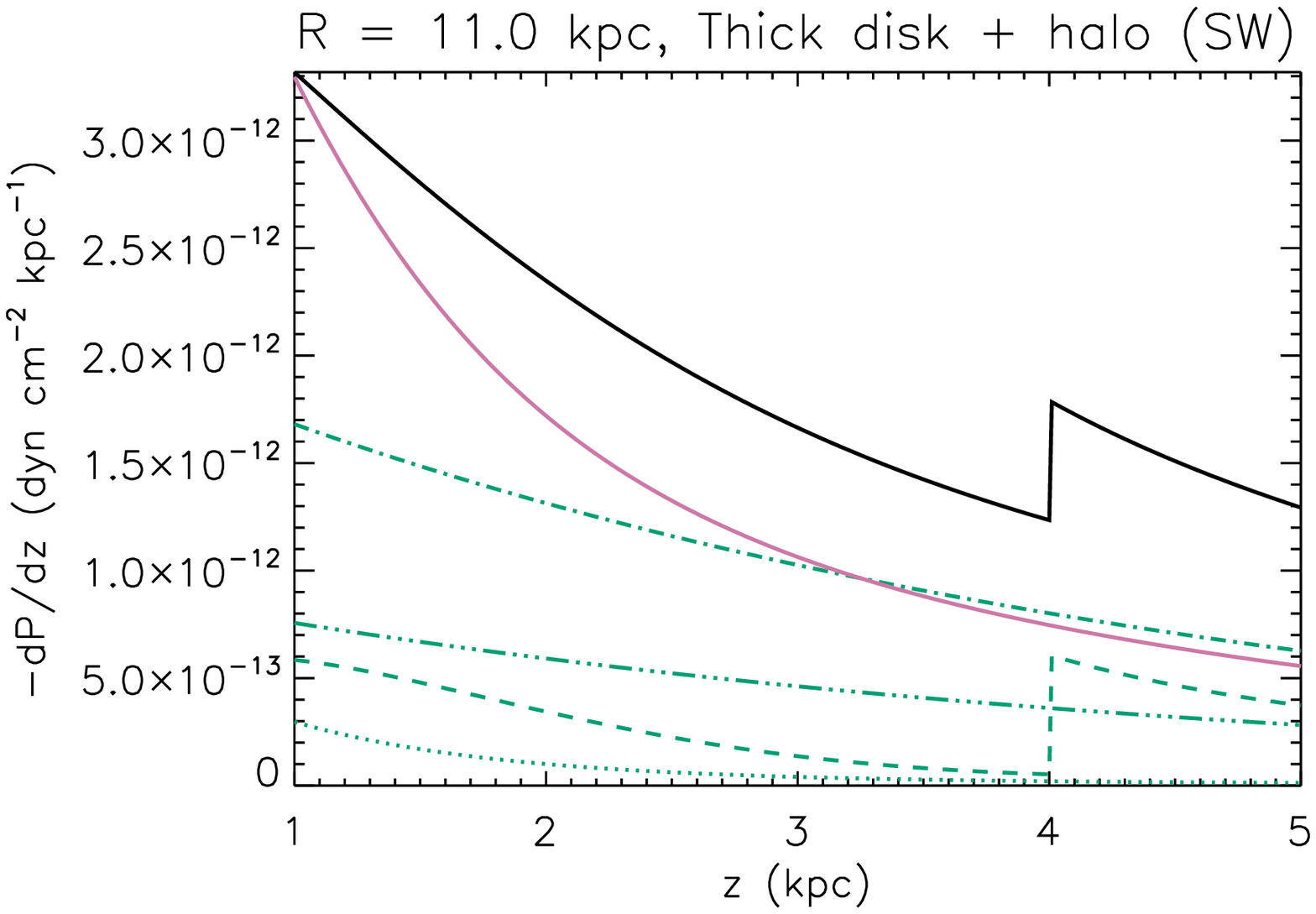}
  \epsscale{1.1}\plottwo{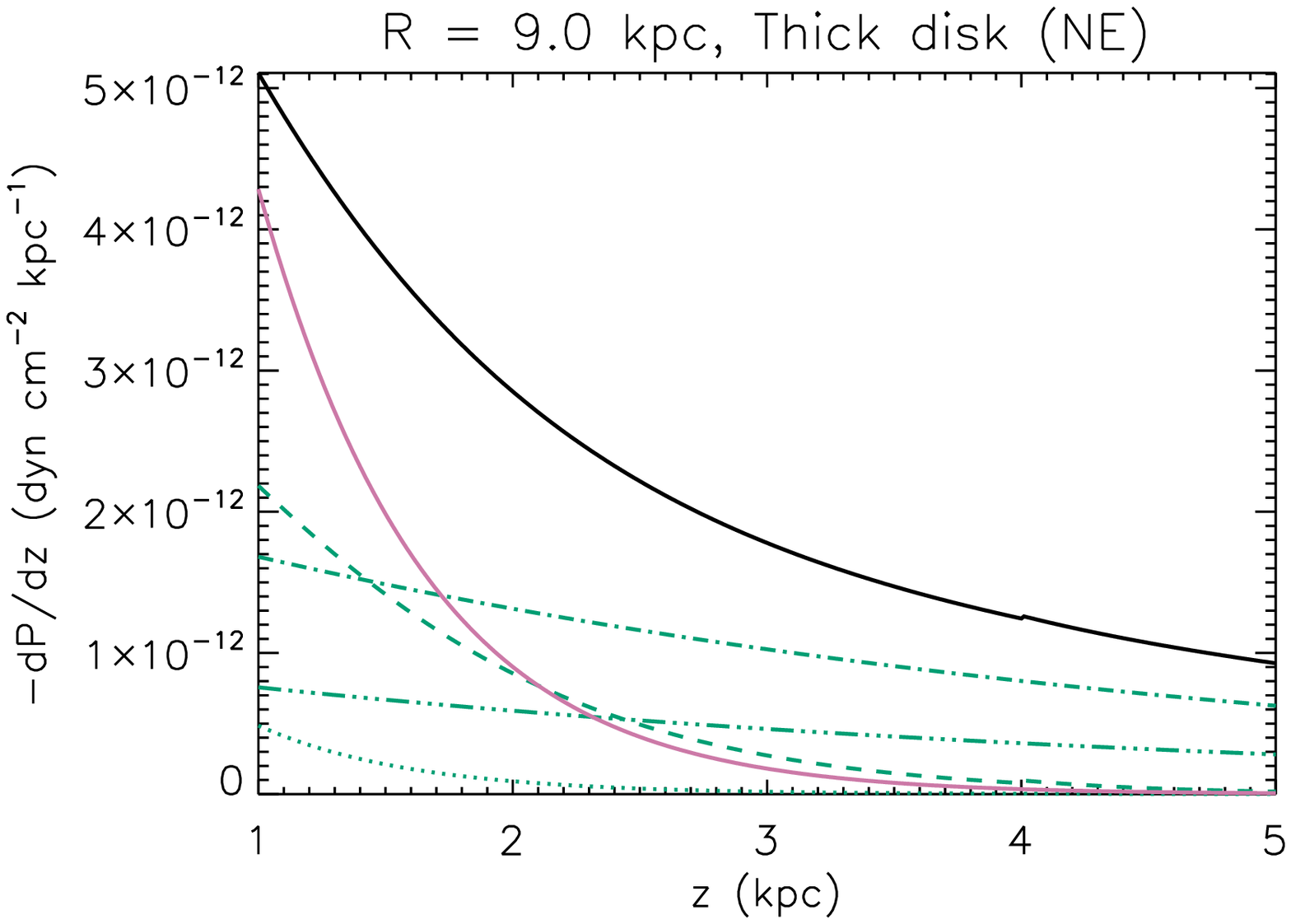}{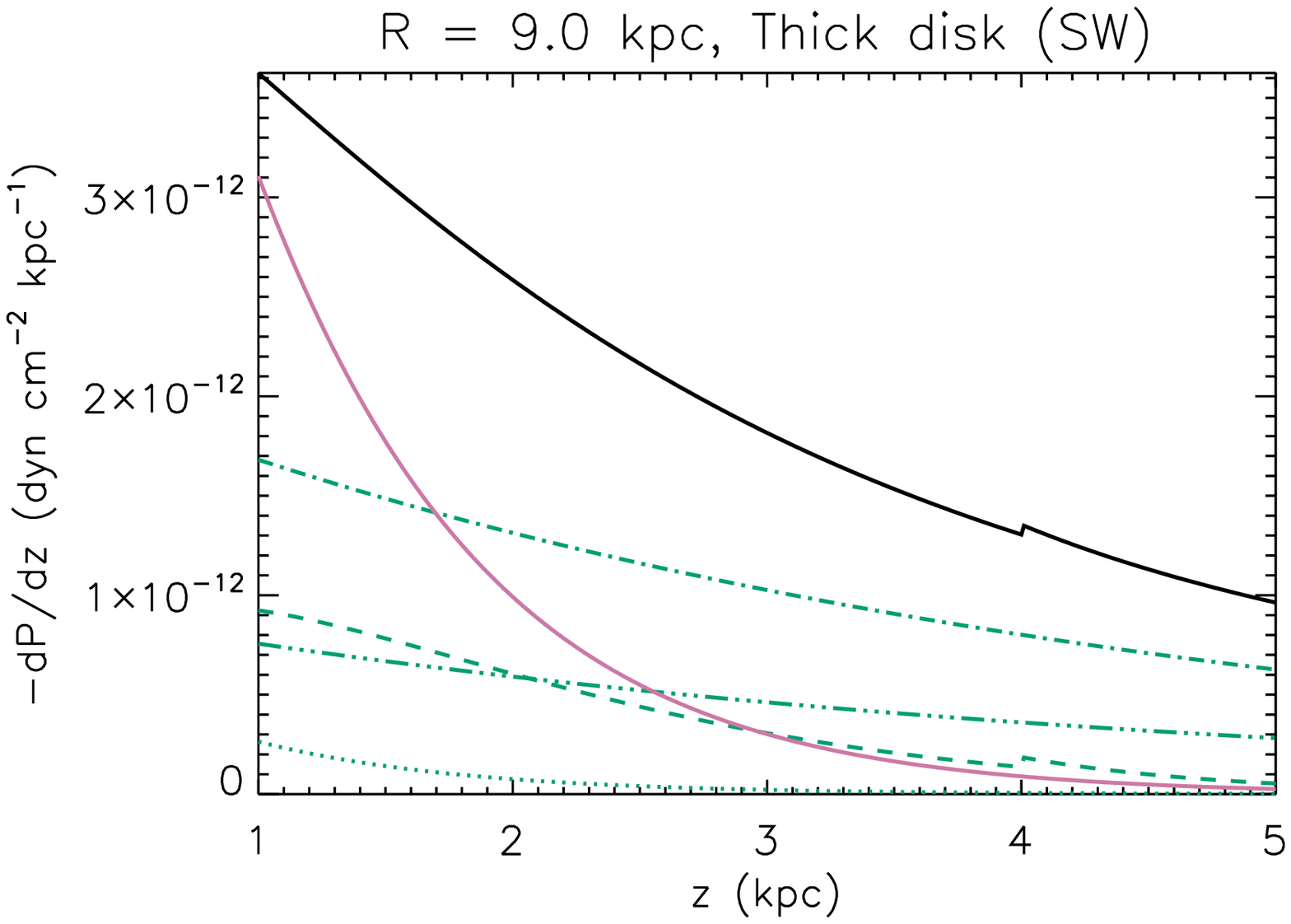}
  \epsscale{1.1}\plottwo{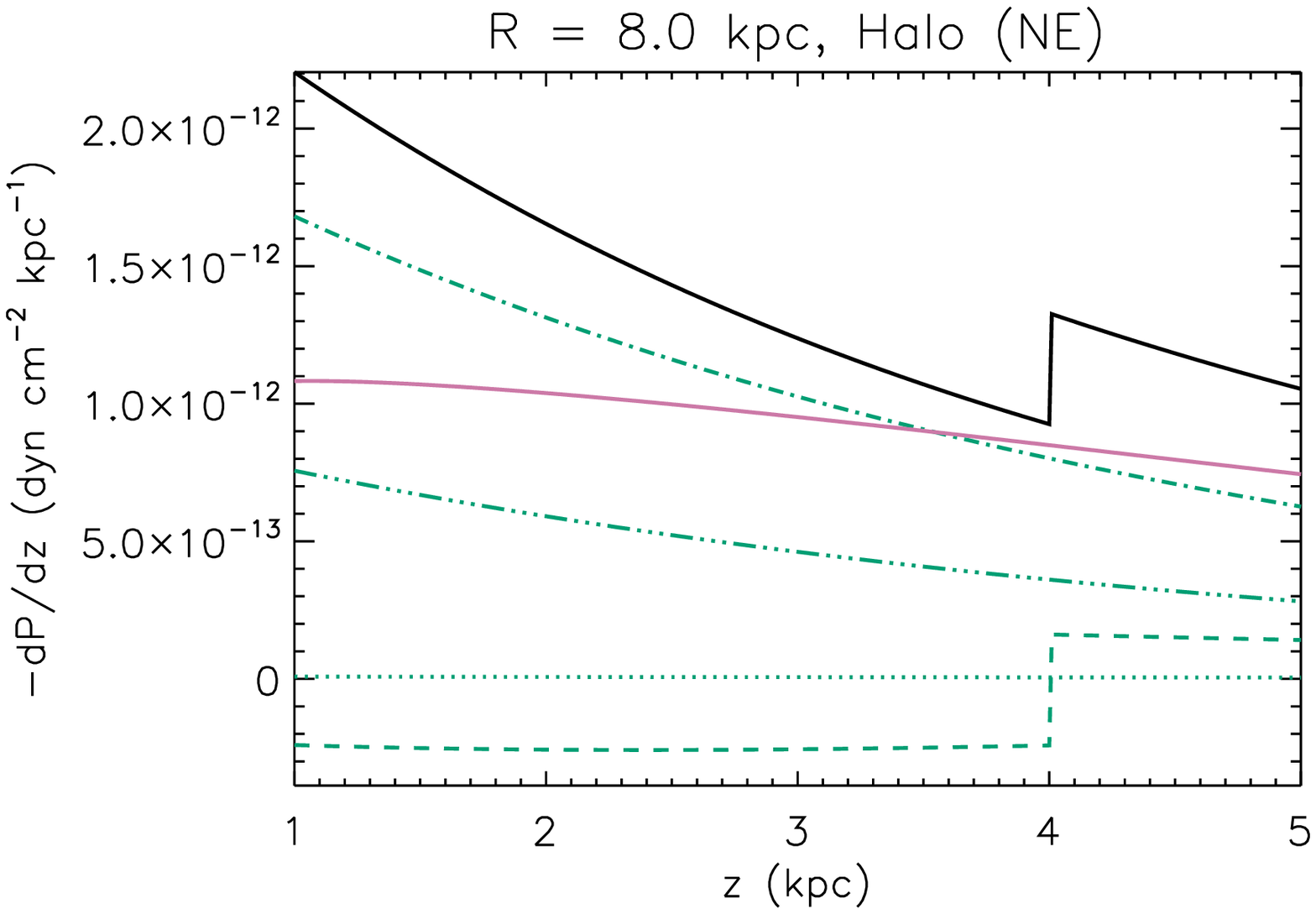}{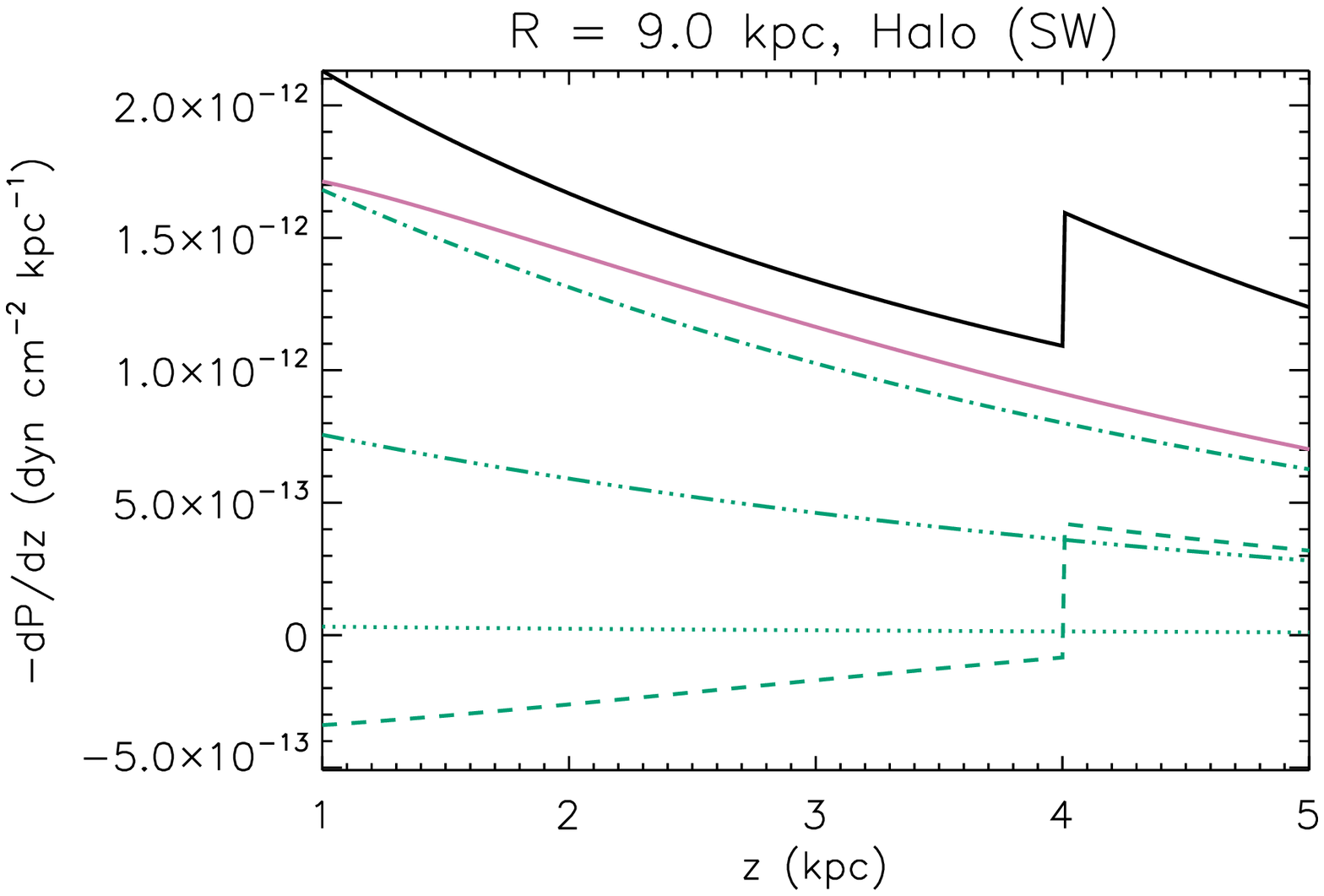}
  \caption{Comparison of observed (black) and required (pink) pressure
    gradients to satisfy a dynamical equilibrium model for the northeast
    (left) and southwest (right) sides of the eDIG layer, shown at the
    \textit{minimum} galactocentric radii at which the dynamical equilibrium
    model is satisfied. At all smaller galactocentric radii, the required
    pressure gradient exceeds the observed gradient, and the model
    fails. Models with thick disk and halo (top), thick disk (middle), and
    halo (bottom) components are considered. The magnetic field is the
    dominant source of vertical pressure support.}
  \label{chap3:fig8}
\end{figure*}

In Fig. \ref{chap3:fig8}, we present the observed and required pressure
gradients at $R_{eq}$ for each side of the galaxy and each eDIG component
considered. The halo components are most easily supported in dynamical
equilibrium, with $R_{eq} = 8$ kpc ($R_{eq} = 9$ kpc) on the northeast
(southwest) sides of the galaxy. Both thick disk components have a similar
result ($R_{eq} = 9$ kpc). Combining the components renders the model more
difficult to satisfy, pushing $R_{eq}$ to $R_{eq} = 10$ kpc on the northeast
side and to $R_{eq} = 11$ kpc on the southwest side.

It is clear that there is insufficient vertical pressure support to satisfy
dynamical equilibrium except in the shallowest parts of the galactic
gravitational potential. In other words, $R_{eq}$ is a significant fraction of
the radial extent of the eDIG layer for all components considered. The success
of such a model thus requires placing the gas in a ring geometry at moderate
to large $R$, perhaps over star-forming spiral arms. It is the magnetic
pressure gradient, followed by the cosmic-ray pressure gradient, that provides
the most substantial source of vertical support.

Note that there is a discontinuity in $\frac{dP_{turb}}{dz}$ at $|z| = 4$ kpc,
where the turbulent velocity dispersion transitions from a rising to a flat
regime. This discontinuity is most prominent in the halo component. At $|z|
\le 4$ kpc, $\frac{dP_{turb}}{dz}$ is positive; this is because
$\sigma(z)^{2}$ rises more quickly than $\rho(z)$ falls due to the large scale
height of the halo. At $|z| > 4$ kpc, $\frac{dP_{turb}}{dz}$ becomes negative
as $\sigma(z)^{2}$ becomes constant and $\rho(z)$ continues to fall. The total
pressure gradient remains negative at all $z$ due to contributions from the
magnetic field and cosmic rays.

From Eq. (\ref{chap3:eq8}), a discontinuity in $\frac{dP}{dz}$ implies a
jump in $\rho(z)$ (in this case, an increase in $\rho(z)$ at $|z| = 4$
kpc). This is an unstable configuration. Thus, we interpret the total observed
pressure gradients in Fig. \ref{chap3:fig8} not as the true pressure
gradients that we seek to reproduce with a density model, but instead as an
indication of the maximum available pressure support in the eDIG layer. This
allows us to assess the viability of the dynamical equilibrium model without
claiming a precise characterization of the true pressure profile. We also note
that the turbulent motions in the halo component may differ from those in the
thick disk; since the latter dominate the line profiles, we do not wish to
overinterpret $\frac{dP_{turb}}{dz}$ in the halo.

We now explore how variations in $\phi$, $B_{0}$, and $h_{z,B}$ affect the
success of the dynamical equilibrium model. We vary $\phi$ from 0.01 (highly
clumpy) to 1 (no clumping) and $B_{0}$ and $h_{z,B}$ both by 50\% ($7.4\ \mu
\text{G} \le B_{0} \le 22.2\ \mu \text{G}$, $4.05\ \text{kpc} \le h_{z,B} \le
12.15\ \text{kpc}$). Within this parameter space, we conduct the analysis
described above for the thick disk and halo combined and determine the value
of $R_{eq}$.

\begin{figure*}[h]
  \epsscale{1.0}\plotone{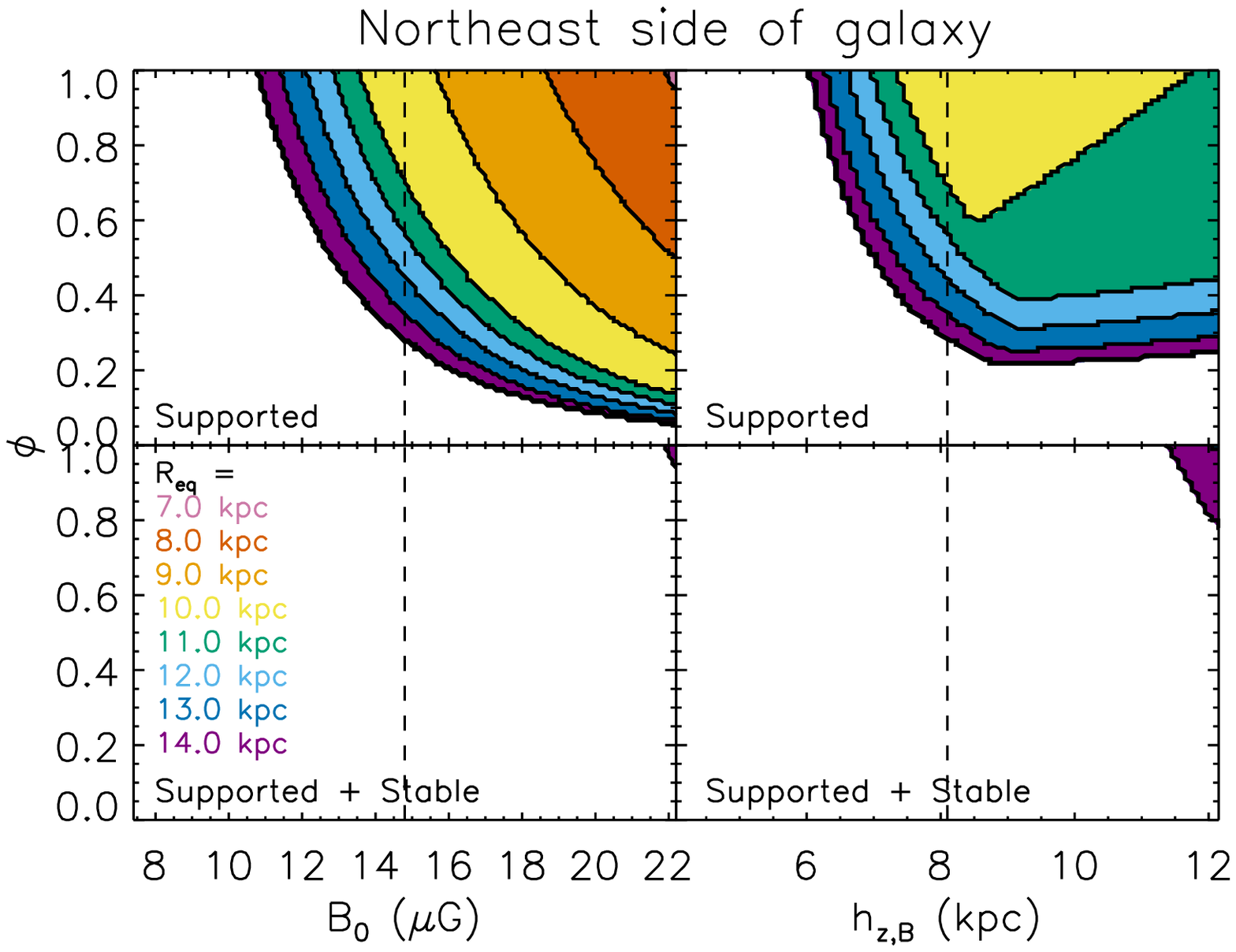} \epsscale{1.0}\plotone{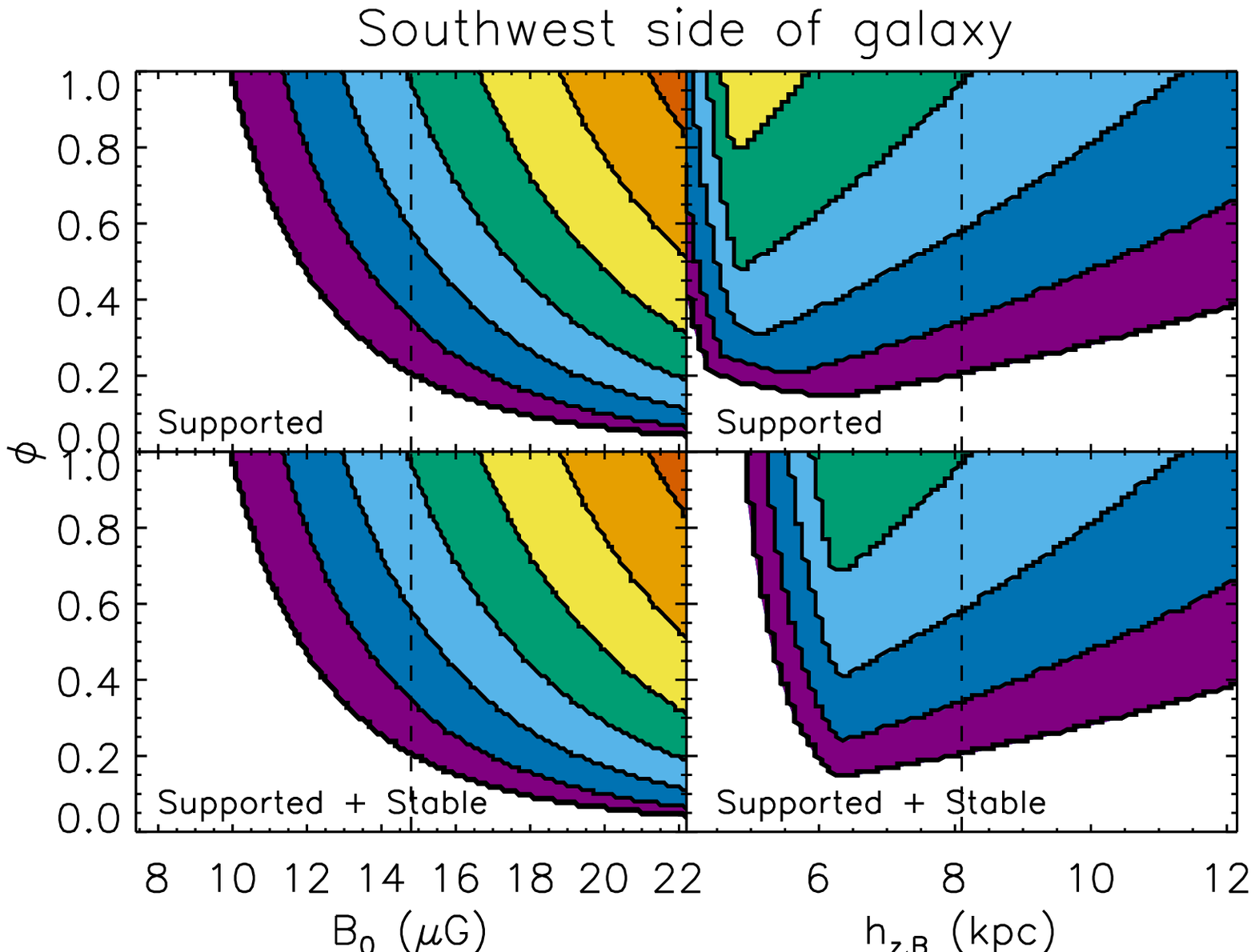} \caption{Exploration
  of the success of a dynamical equilibrium model for the eDIG layer
  over a range of $\phi$, $B_{0}$, and $h_{z,B}$. Both the thick disk
  and halo components of the eDIG layer are included in the model for
  the northeast (top) and southwest (bottom) sides of the galaxy. We
  plot the minimum galactocentric radius at which the dynamical
  equilibrium model succeeds, $R_{eq}$, in the top panels of both
  figures (labeled ``Supported''). We also require the stability
  criterion given in Eq. \ref{chap3:eq17} to be satisfied in the
  bottom panels (labeled ``Supported and Stable''). The large values
  of $R_{eq}$ required for a stable, successful dynamical equilibrium
  model on the northeast side effectively eliminates this model at
  this location. The dashed lines indicate the preferred values of
  $B_{0}$ and $h_{z,B}$. In the lefthand (righthand) column, $h_{z,B}
  = 8.1$ kpc ($B_{0} = 14.8\ \mu$G).}  \label{chap3:fig9}
\end{figure*}

We show results in this parameter space in Fig. \ref{chap3:fig9} for
the northeast (top) and southwest (bottom) sides of the galaxy. In the
top two panels for each side, we display $R_{eq}$ in the $\phi$,
$B_{0}$ ($h_{z,B} = 8.1$ kpc) and $\phi$, $h_{z,B}$ ($B_{0} =
14.8\ \mu$G) planes. Everywhere in these planes, $R_{eq} \ge 7$ kpc;
this confirms the need for a ring geometry for the dynamical
equilibrium model to succeed.

In general, for the gas to be supported at a given galactocentric radius, a
higher magnetic field strength (a steeper $\frac{dP_{B}}{dz}$) is required for
a smaller $\phi$ (a steeper $\frac{d\rho}{dz}$). Likewise, a smaller $h_{z,B}$
(a steeper $\frac{dP_{B}}{dz}$) is needed to support the gas given a smaller
$\phi$. As the eDIG scale height begins to exceed the magnetic scale height,
it is more difficult to support the eDIG at $|z| > h_{z,B}$, and larger values
of $\phi$ are again required.

We briefly consider the impact of a radially-dependent eDIG density
distribution on the success of the dynamical equilibrium model. For
radial dependence of the form $n_{e}(R) \propto e^{-R/h_{R}}$, a
reasonable choice of $h_{R}$ (i.e., a few kpc, comparable to the
stellar scale length) has only a small effect on the model's
success. For example, a choice of $h_{R} = 4$ kpc mildly reduces
$R_{eq}$ to 8 kpc from 10 and 11 kpc for the combined thick disk and
halo models on the northeast and southwest sides of the disk,
respectively. This minor reduction in $R_{eq}$ is expected if the gas
surface density decreases with $R$, rendering the model easier to
satisfy at large galactocentric radii. In the absence of a comparable
radial dependence in the magnetic field strength, such a radial
fall-off in the density distribution produces a magnetic pressure
comparable to the gas pressure (thermal and turbulent) at small $R$,
but in excess of the gas pressure by an order of magnitude at $R = 10$
kpc. In a model lacking radial dependence, the magnetic pressure
exceeds the gas pressure by a factor of $\sim 2$, more consistent with
approximate energy equipartition.

\subsection{Stability Analysis}
\label{chap3:sec:stab}

The magnetic buoyancy instability first described by \citet{Parker1966}
presents another challenge to the success of our dynamical equilibrium
model. Although Parker considered cosmic rays as a fluid with an adiabatic
index $\gamma_{cr} = 0$, we allow for cosmic ray coupling to the gas either
via extrinsic turbulence or via the streaming instability for high and low
energy cosmic rays, respectively ($\gamma_{cr} = 1.45$ given $P_{cr} =
0.45U_{cr}$; \citealt{Zweibel2013}). This allows the cosmic rays to act toward
stabilizing the gas as well as elevating it above its natural scale height.

These considerations yield the following stability criterion
\citep{Newcomb1961, Parker1966, Zweibel1975}:
\begin{equation}
  -\frac{d\rho}{dz} > \frac{\rho^{2}g_{z}}{\gamma_{g}P_{g} +
    \gamma_{cr}P_{cr}}.
  \label{chap3:eq17}
\end{equation}
To determine whether the eDIG layer is robust against the instability, we
calculate the minimum value of $\gamma_{g}$ required to satisfy this criterion
for both $\gamma_{cr} = 0$ (no cosmic ray coupling) and $\gamma_{cr} = 1.45$
(cosmic ray coupling). An adiabatic index of $\gamma_{g} = 5/3 - 2$ may be
characteristic of the star-forming, turbulent ISM \citep{Zweibel1975};
however, values as low as $\gamma_{g} = 1$ may be found in the eDIG
\citep{Parker1966}. Therefore, we define a stable model as that for which the
minimum value of $\gamma_{g}$ that satisfies the stability criterion is
$\gamma_{g} \le 1$.

\begin{figure*}[h]
  \epsscale{1.1}\plottwo{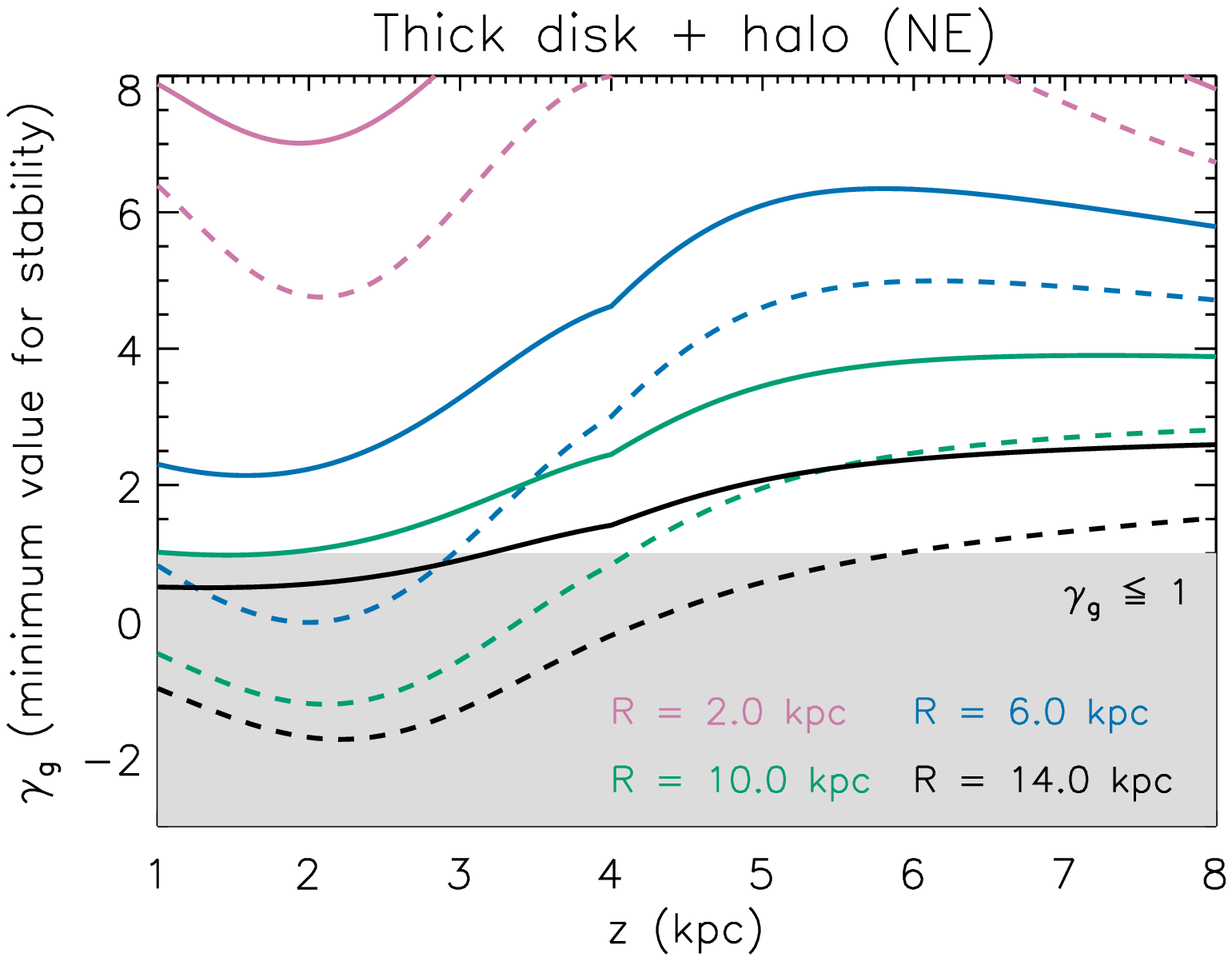}{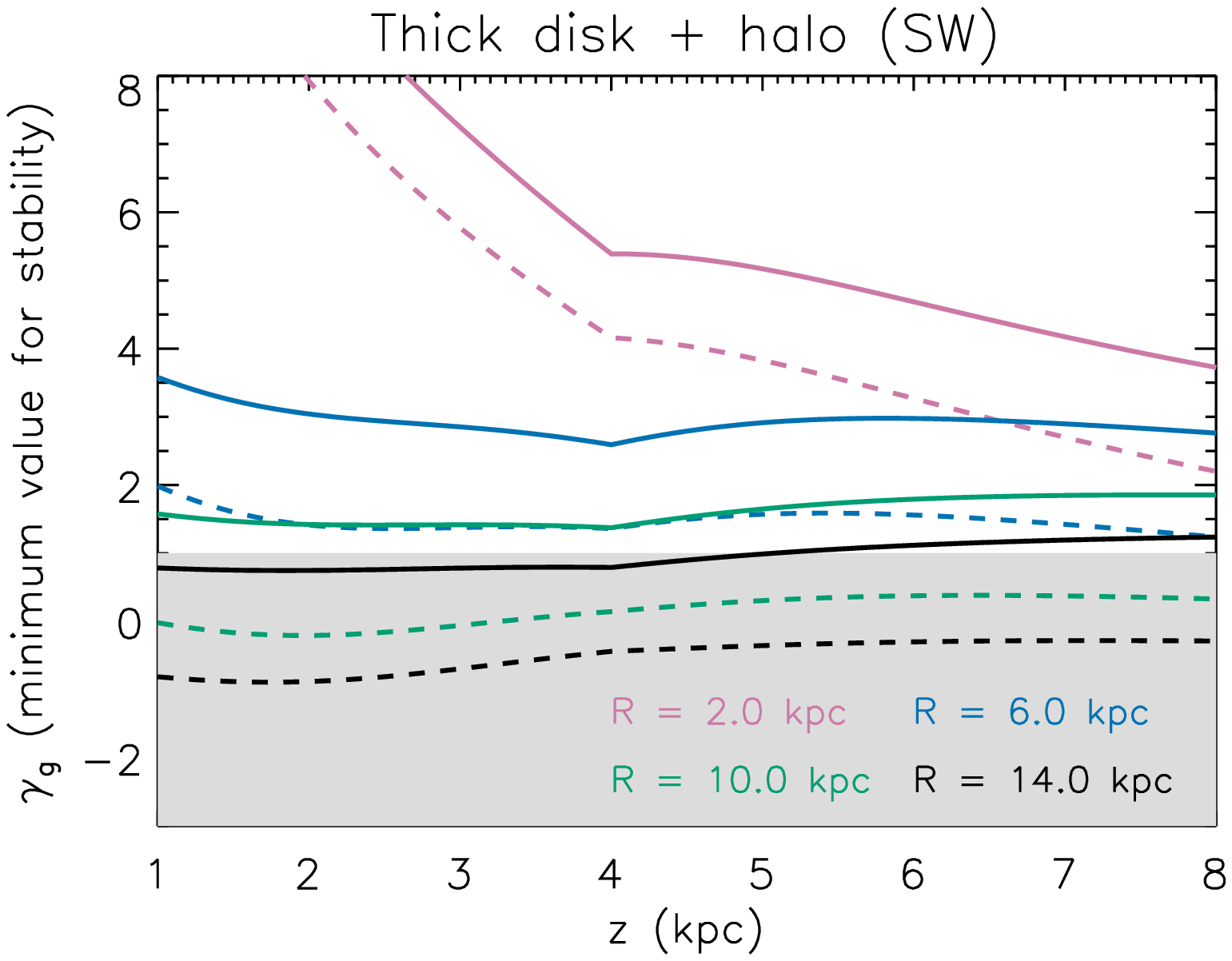}
  \epsscale{1.1}\plottwo{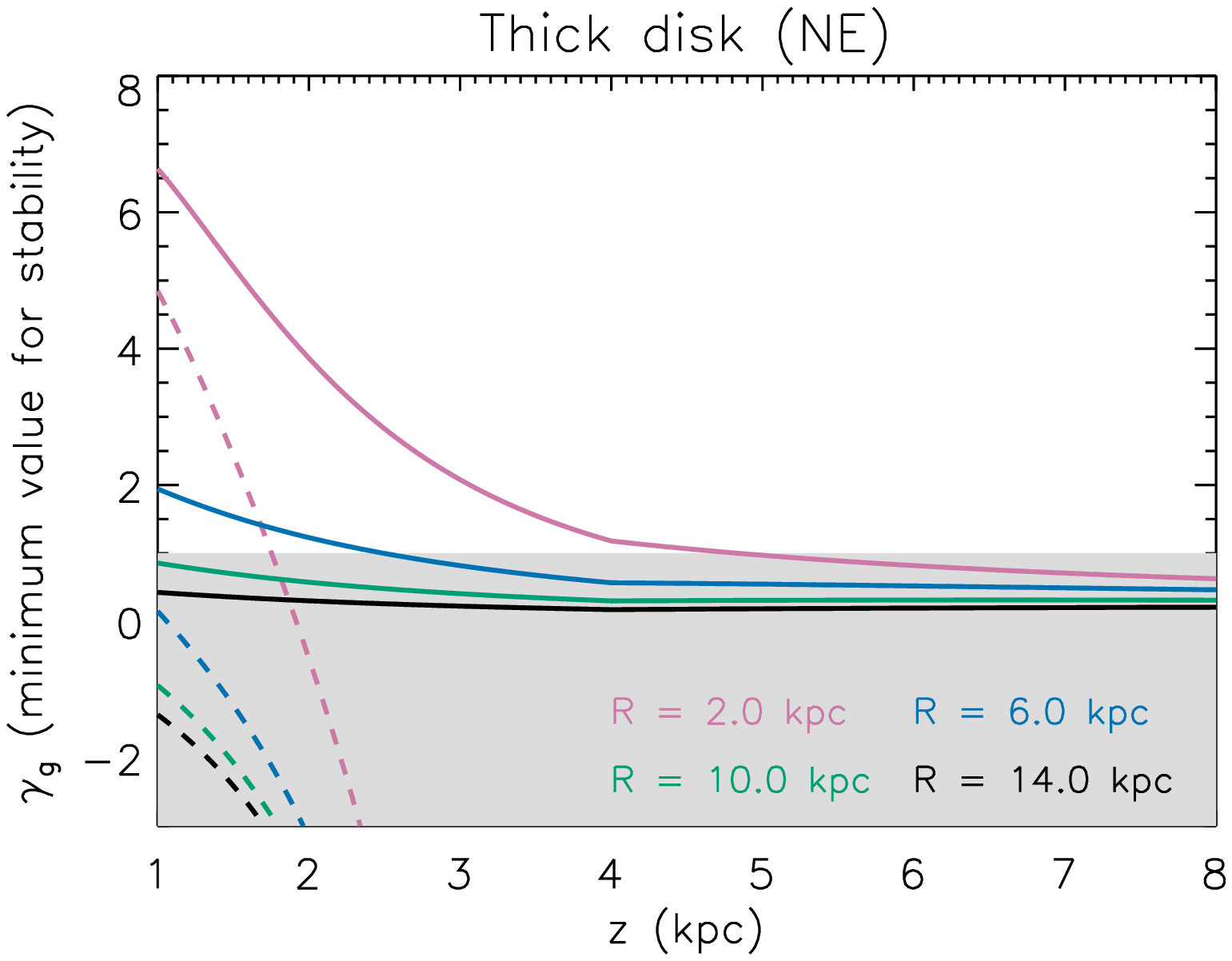}{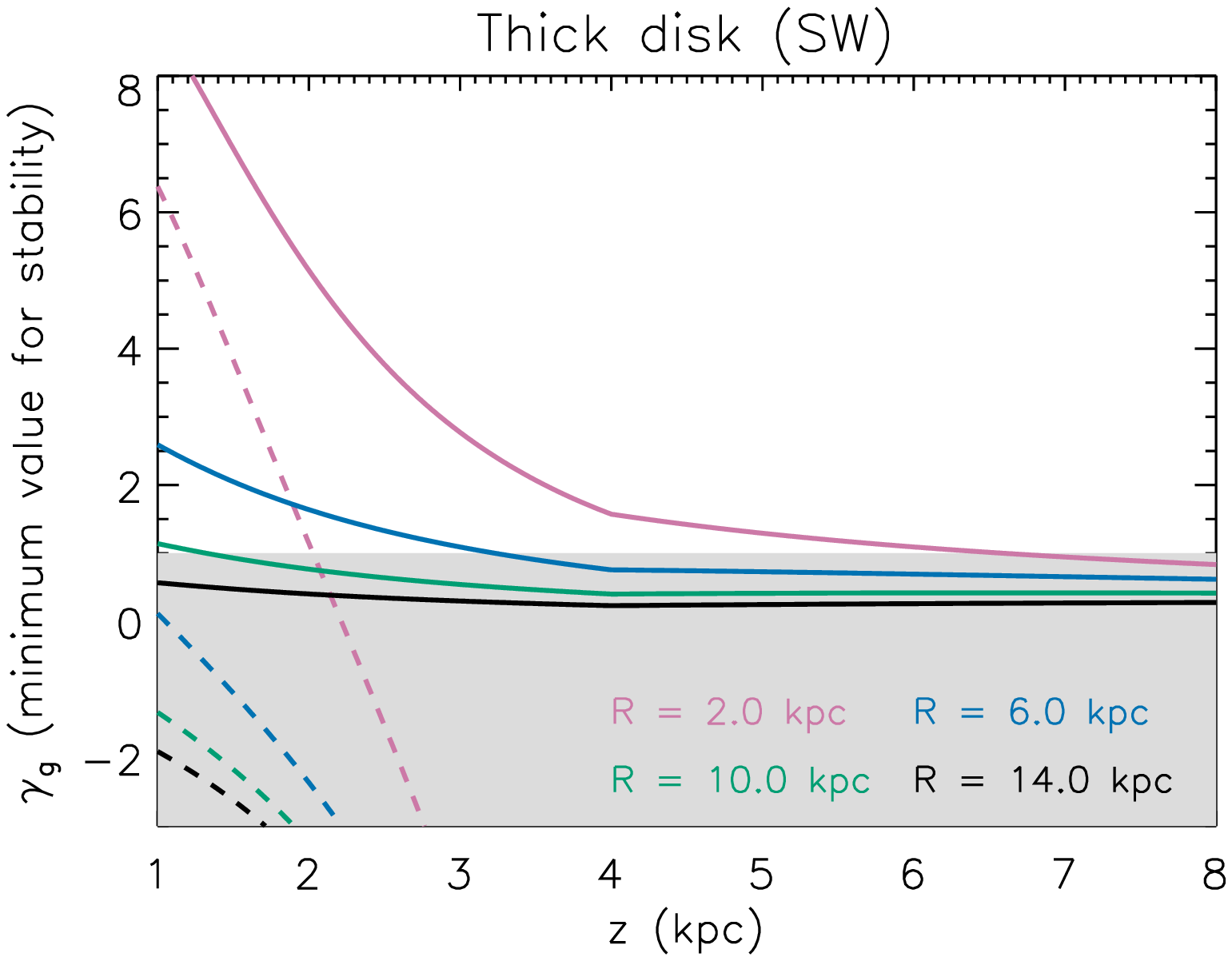}
  \epsscale{1.1}\plottwo{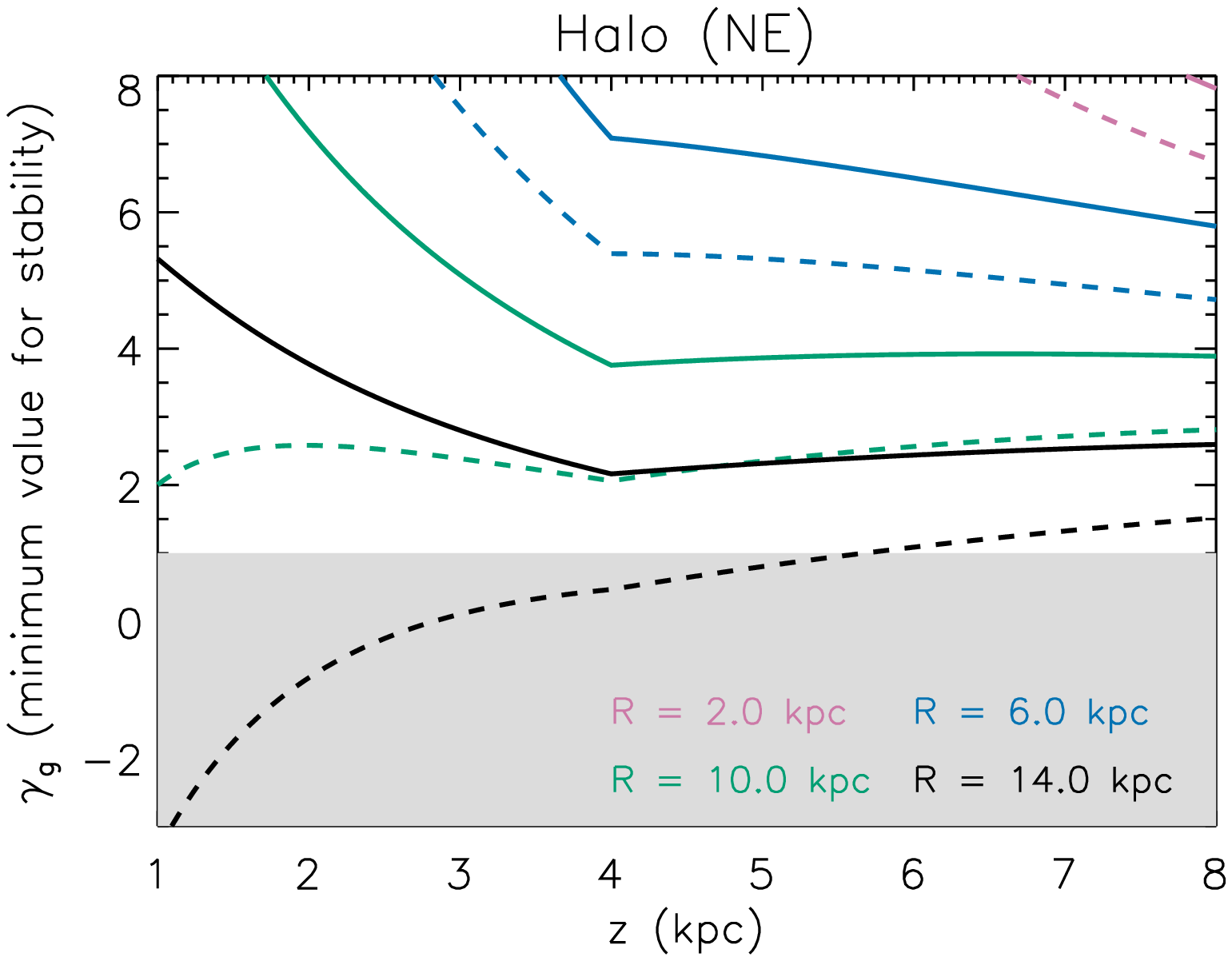}{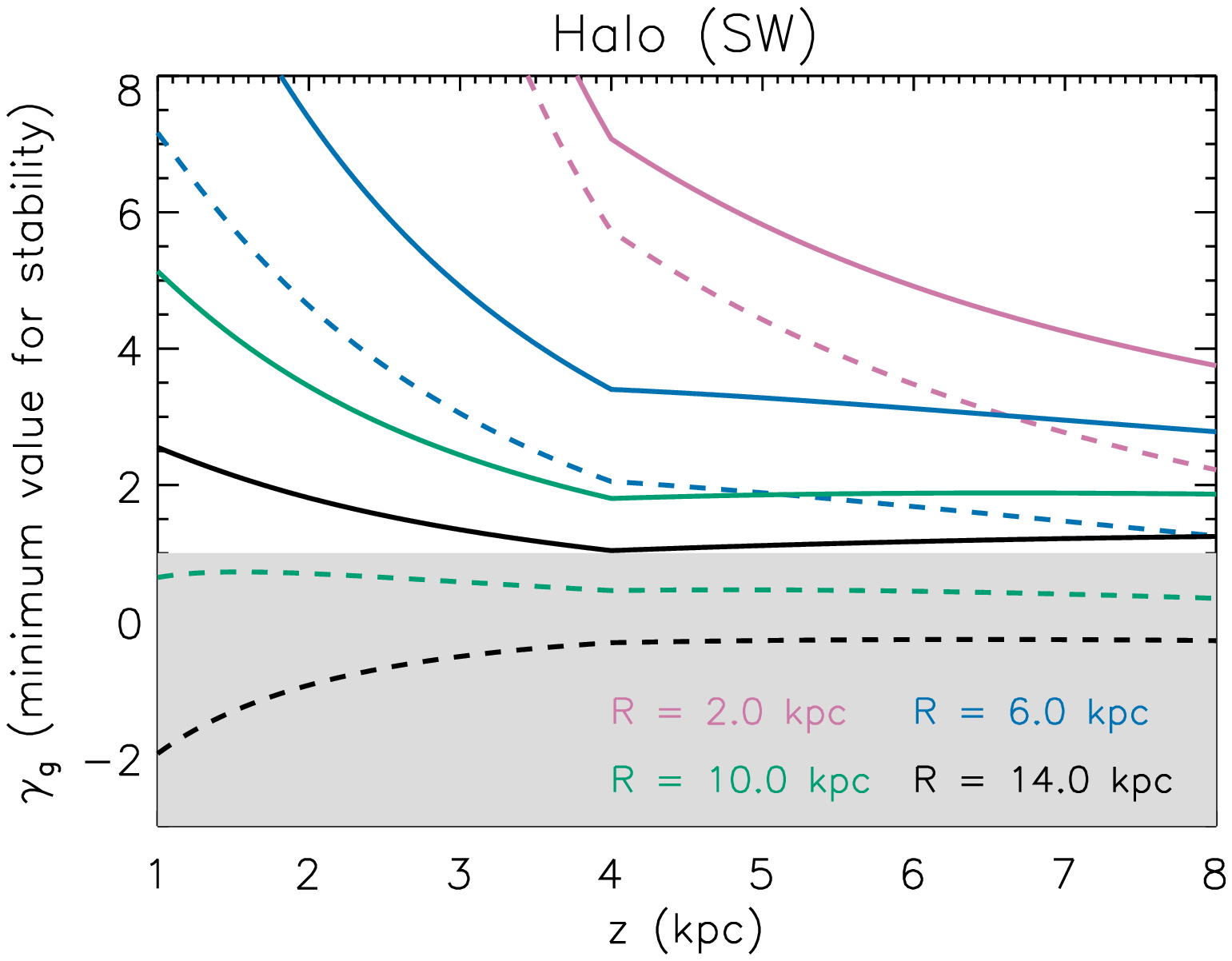}
  \caption{Minimum value of $\gamma_{g}$ required to satisfy the Parker
    stability criterion for the eDIG layer at a range of
    galactocentric radii. Solid and dashed lines are derived assuming
    $\gamma_{cr} = 0$ and $\gamma_{cr} = 1.45$. Models within the shaded
    region satisfy the conservative criterion that the minimum $\gamma_{g}$
    required for stability be $\gamma_{g} \le 1$. This requirement is only
    satisfied at all $z$ for $\gamma_{cr} = 1.45$, and even then only at large
    galactocentric radii (generally $R \ge 6$ kpc). The Parker instability
    thus further restricts the success of the dynamical equilibrium model to
    moderate and large galactocentric radii.}
  \label{chap3:fig10}
\end{figure*}

\citet{Heintz2018} have recently demonstrated the potential importance
of cosmic-ray streaming to the stability of the magnetized ISM. They
illustrate that inclusion of cosmic-ray streaming can be destabilizing due to
cosmic-ray heating. This effect is mitigated in the isothermal case
($\gamma_{g} = 1$). Future work will explore the implementation of cooling and
the characteristic wavelength of the instability in the nonlinear regime,
potentially shedding light on the impact, if any, of cosmic-ray heating on the
electron temperature and/or ionization structure of the eDIG. However, it is
important to note that even if a model satisfies the stability criterion
above, the model may be unstable if heating via cosmic-ray streaming is
important.

We perform the stability analysis between $1\ \text{kpc} \le |z| \le 10\
\text{kpc}$. It is important to note that our treatment of the instability is
based on a one-dimensional model and may fail at large $z$. In Fig.
\ref{chap3:fig10}, we present the minimum values of $\gamma_{g}$ required to
satisfy the stability criterion for the thick disk and the halo, separately
and combined. Models with $\gamma_{cr} = 0$ are not stable at all $z$ for the
relevant $R$; thus, it is clear that the cosmic rays must be coupled to the
gas to achieve a stable model.

Allowing $\gamma_{cr} = 1.45$, the thick disk is stable at all radii at which
the dynamical equilibrium model succeeds. In contrast, the halo on the
northeast (southwest) side of the disk is not stable at all $z$ until $R > 14$
kpc ($R \ge 10$ kpc). The instability of the halo is exacerbated by the large
scale height on the northeast side. A gas cloud perturbed upwards will be
buoyant if the rate at which its internal density decreases is faster than the
rate at which the background density decreases. This is more likely to occur
in a medium in which the gas density profile changes only slowly, rendering
the halo less stable than the thick disk. While the coupled cosmic rays can
``stiffen'' the equation of state of the medium and thus decrease the minimum
galactocentric radius at which a stable model can be achieved, this effect is
insufficient to produce stable models outside of a thin ring, or at all, on
the southwest and northeast sides of the disk, respectively.

We now consider the effect of stability on the parameter space study discussed
in the previous section. We ask where in parameter space a \textit{stable},
successful dynamical equilibrium model is found, and show the results in
Fig. \ref{chap3:fig9}. For the northeast side of the disk, $R_{eq}$
increases to $R_{eq} \geq 14$ kpc as compared to $R_{eq} \geq 7$ kpc in the
slices of parameter space previously considered; this is due to the
instability of the halo at large $z$ seen in Fig. \ref{chap3:fig10}. On the
southwest side of the disk, for which the halo scale height is a factor of
$\sim 2$ smaller than on the northeast side, the areas of parameter space over
which the dynamical equilibrium model succeeds and over which it succeeds
\textit{and} is stable are much more comparable. The instability of the
dynamical equilibrium model at almost all galactocentric radii on the
northeast side effectively rules out this model for this side of the galaxy.

\section{Discussion}
\label{chap3:sec:disc}

Comparing the results of \citet{Boettcher2016} to those presented here, a
dynamical equilibrium model fails in a similar way for the eDIG layers in the
Milky-Way analog NGC 891 and the interacting, star-forming galaxy NGC 5775. In
both systems, stably supporting the eDIG layer in dynamical equilibrium is
only feasible at moderate to large galactocentric radii, requiring a contrived
ring geometry that is inconsistent with observations of face-on disk galaxies
\citep[e.g.,][]{Boettcher2017}. Such a model relies predominantly on the
magnetic field and cosmic ray pressure gradients to support the eDIG
layers at their observed scale heights.

Although the great vertical extent of the eDIG layer in NGC 5775 may
follow from its enhanced star-formation rate per unit area, factor(s)
other than the star-formation rate surface density set the vertical
extents of the synchrotron halos in the CHANG-ES sample analyzed
by \citet{Krause2018}. In this sample, the synchrotron scale heights
are positively correlated with the synchrotron diameters of the
galaxies. Although more work is needed to interpret this trend in the
context of eDIG layers, it appears that the processes that govern the
spatial extents and pressure profiles of the thermal and nonthermal
halos do not conspire together to achieve an equilibrium configuration
at the disk-halo interface.

We note that there are several caveats to the dynamical equilibrium model
presented here. First, we assume that the measured horizontal velocity
dispersion is equal to the vertical velocity dispersion. If the vertical
dispersion exceeds the horizontal dispersion, as suggested by observations of
M83 \citep{Boettcher2017}, then we would underestimate the turbulent pressure
support here. Additionally, we do not consider the potential roles of magnetic
tension \citep{Boulares1990}, radiation pressure \citep[e.g.,][]{Franco1991},
and the hot halo gas in supporting the eDIG layer at its observed scale
height.

The role of the hot halo is of particular interest due to the detection of
diffuse, spatially-extended, X-ray-emitting gas in the halo of NGC 5775
\citep{Li2008, Li2013}. If the warm and hot phases are sufficiently coupled,
then the vertical pressure gradient in the hot halo may merit inclusion in the
pressure gradients considered here. This is especially true given the
likelihood of a large volume filling factor for the hot phase ($\phi_{hot}
\sim 1$). Thus, we make a simple estimate of the vertical pressure support
provided by the hot halo.

\cite{Li2008} find two temperature components in the hot, diffuse phase, with
electron densities of a few $\times 10^{-3}\ \text{cm}^{-3}$ and temperatures
of a few $\times 10^{6}$ K. This corresponds to a thermal velocity dispersion
of $\sigma \ge 200$ \kms and thus a thermal scale height of $h_{z} \sim 10$
kpc at a galactocentric radius of $R \sim 5$ kpc in the galactic gravitational
potential given in the Appendix. Assuming a hot halo in hydrostatic
equilibrium, the vertical thermal pressure gradient in the hot phase is
comparable to that in the warm phase. As we saw in Fig. \ref{chap3:fig8}, the
thermal pressure gradients are small compared to the nonthermal pressure
gradients, and thus the hot phase does not appear to substantially improve the
performance of the dynamical equilibrium model.

If the hot phase has a scale height considerably less than its thermal scale
height, then it may be able to contribute more substantially to the pressure
support. This could occur if the hot phase is confined by a magnetic field
that is anchored to the disk via the warm medium. The relatively small scale
height observed by \citet{Li2008} for the hot phase ($h_{z} \sim 1.5$ kpc)
suggests that such a scenario may be appropriate. However, in this case, a
warm and hot phase with comparable scale heights in pressure equilibrium can
at most double the contribution to the thermal pressure gradient compared to
the warm medium alone. As we have seen, this remains insufficient to support
the layer, and thus we conclude that the hot phase is not the source of the
missing pressure support in a dynamical equilibrium context. Note that in
addition to a diffuse background, a superbubble is detected in X-ray emission
to distances of $|z| \sim 10$ kpc on the southern side of the galaxy. An
expanding or outflowing hot halo may impart momentum to the eDIG and thus
enhance its scale height, but we consider this as a separate class of models
than the hydrostatic case tested here.

Additional assumptions may have affected the success of the dynamical
equilibrium model. We employ a simple, plane-parallel magnetic field geometry
despite evidence that the field becomes X-shaped at large distances from the
midplane \citep{Soida2011}. An increasingly vertical component of the magnetic
field with height above the disk introduces the possibility that magnetic
tension plays a role in supporting the gas, an effect that appears important
in the Galaxy \citep{Boulares1990}. The presence of a thin synchrotron disk
also suggests a magnetic field component with a scale height significantly
smaller than the eDIG scale height(s), a potentially unstable configuration in
which the gas layer would ``float'' atop the magnetic field.

We determine the magnetic field and cosmic-ray pressure gradients via the
energy equipartition assumption despite increasing indications for departures
from equipartition in a range of environments \citep[e.g.,][]{Schmidt2016,
  YoastHull2016}. We also must make simplifying assumptions about the
cylindrical geometry and radial extent of the eDIG layer that affect our
estimates of the electron density profile. While adjustments to any of these
assumptions could affect the success of the dynamical equilibrium model, its
failure in two galaxies with disparate properties builds confidence that this
model is a poor description of the dynamical state of eDIG layers.

Given the failure of the dynamical equilibrium model, we can ask what
evidence exists for nonequilibrium models of the gaseous, disk-halo
interface. The eDIG emission-line ratios suggest that the gas
is chemically enriched. This implies that the eDIG is either directly
transported out of the disk or originates in a disk-halo circulation
that triggers thermal instabilities in the lower
halo. Chemically-enriched, warm gas may arise through cooling in the
wakes of cool clouds traveling through the hot halo if either the halo
itself is enriched or the clouds sufficiently enrich their
wakes \citep[see, e.g.,][]{Marasco2012}. Careful study of chemical
abundances inferred from emission-line ratios may shed light on the
extent to which such ``induced'' accretion contributes to the eDIG
layer in NGC 5775. Regardless, the evidence suggests that the
low-angular-momentum gas originates in a disk-halo flow rather than
through direct accretion from a metal-poor, intergalactic reservoir.

We discuss the two most commonly considered disk-halo flows - galactic
fountains and galactic winds - here. We favor the former scenario for
several reasons. First, we do not see evidence for a traditional
outflow cone in the eDIG kinematics. If such a cone exists, it must
have an unusually small opening angle that directs the outflow away
from our line of sight. Although the symmetric, blueshifted gas at
large $z$ along the minor axis hints at an outflow, it lacks the
redshifted component that we expect to arise from an outflow cone and
has a line-of-sight velocity well below the escape velocity. If an
outflow is present in the hot phase of the halo, as suggested by the
X-shaped field geometry \citep{Soida2011}, then a small volume filling
factor would likely be required for the eDIG layer to avoid
entrainment of the warm gas.

Thus, we favor a galactic fountain flow in which eDIG clouds circulate
between the disk and the halo. The blueshifted gas at large $z$ along
the minor axis may be evidence of such a circulation. We may only
observe the outflow stage of the fountain flow if the gas is primarily
in a warm, ionized phase during this stage. However, this is somewhat
at odds with the conventional picture in which gas clouds
leave the disk in a hot ionized phase and return to the midplane in a
warm ionized or neutral state. An alternative explanation is that the
kinematics observed along the minor axis are dominated by a local
feature such as a filament that is primarily outflowing from the
disk. However, no such filament is obvious in the H$\alpha$ imaging of
\citet{Collins2000}.

We also consider whether the interaction with NGC 5774 has influenced the
kinematics of extraplanar gas in NGC 5775. There is no obvious evidence of
interaction in any of the slit locations; most notably, s2 produces a rotation
curve characteristic of noninteracting galaxies despite sampling the side of
the galaxy where the HI bridge connects the companion galaxies. However, the
question of whether the blueshifted gas along the minor axis could arise from
ram pressure is intriguing. If the interaction has deposited comparatively
dense gas in the path of NGC 5775, then passage through this gas could sweep
the halo of NGC 5775 toward the observer as the galaxy recedes. Although the
HI kinematics of the NGC 5775 - NGC 5774 system are potentially consistent
with this scenario \citep{Irwin1994}, more detailed modeling is needed to
assess the effects of ram pressure on the gaseous halo. We also note that such
effects are not necessary to interpret the kinematics observed off of the
minor axis in this work and others \citep[e.g.,][]{Rand2000}.

It is clear that the current observations do not provide a fully definitive
description of the dynamical state of the eDIG layer in NGC 5775. However,
they suggest a most likely scenario. The interaction of NGC 5775 with NGC
5774, while not directly shaping the kinematics of extraplanar gas in the
inner disk, has enhanced the star-formation rate in the former galaxy and thus
helped give rise to the vertically-extended, multiphase gaseous halo. The
warm, ionized phase has a multicomponent vertical density distribution that
is most likely sustained via a galactic fountain flow between the disk and the
halo. The observed increase in the velocity dispersion as a function of height
is a natural consequence of this model if the gas clouds are ejected from the
disk with a range of velocities; in this model, the clouds with the highest
dispersion will naturally reach the largest scale heights. The evidence for
shock heating and ionization in the emission-line ratios is consistent with
this turbulent, multiphase medium. We discuss future observations to explore
this interpretation in the conclusions below.

\section{Summary and Conclusions}
\label{chap3:sec:summ}

We combined optical and NUV emission-line spectroscopy from RSS on SALT with
radio continuum observations from CHANG-ES to study the dynamical state of the
eDIG layer in NGC 5775. We summarize the main results of our study here:
\begin{itemize}
\item The exponential electron density distribution has both thick disk and
  halo components and is asymmetric on the northeast and southwest sides of
  the galaxy. The scale heights of these components are $h_{z,e} = 0.6\
  \text{and}\ 7.5$ kpc on the northeast and $h_{z,e} = 0.8\ \text{and}\ 3.6$
  kpc on the southwest sides, showing the remarkable spatial extent of the
  warm ionized halo.
\item We observe evidence of a previously-detected rotational velocity
  gradient characteristic of lagging halos \citep[e.g.,][]{Heald2006a}. We
  report line-of-sight velocity gradients as high as $\Delta v = -25\
  \text{\kms}\ \text{kpc}^{-1}$ in some locations, although quantification of
  the true \textit{rotational} velocity gradient is beyond the scope of this
  work.
\item The H$\alpha$ velocity dispersion displays a significant increase from
  $\sigma_{H\alpha} = 20$ \kms at $z = 0$ kpc to $\sigma_{H\alpha} = 60$ \kms
  at $|z| = 4$ kpc, where the dispersion appears to plateau. This is the first
  clear evidence for a rising velocity dispersion as a function of height in
  an eDIG layer.
\item The thermal, turbulent, magnetic field, and cosmic-ray pressure
  gradients are insufficient to support the gas in dynamical equilibrium at
  its observed scale height except at large galactocentric radii ($R \ge 11$
  kpc for the sum of all density components). Success of this model thus
  requires a ring geometry inconsistent with observations of face-on galaxies.
\item In the context of the dynamical equilibrium model, we assess the
  stability of the eDIG layer against the well-known Parker instability. While
  the thick-disk component is largely stable if the cosmic rays are coupled to
  the gas ($\gamma_{cr} = 1.45$), the halo component on the northeast side is
  not stable at all $z$ over the range of galactocentric radii considered ($R
  \le 14$ kpc).
\item The emission-line ratios display smooth trends from the disk to the halo
  consistent with increased heating of uniform-metallicity gas at large
  $z$. This suggests a disk origin for the gas; if the eDIG is evidence of
  accretion activity, it must arise from an enriched halo.
\item There is no obvious evidence of interaction between NGC 5775 and NGC
  5774 in the kinematics of the extraplanar gas. It appears that the main
  influence of the interaction on the thick-disk ISM of NGC 5775 is to trigger
  the star formation that gives rise to the multiphase, gaseous halo.
\item At large $z$ along the minor axis, the gas is blueshifted on both sides of the midplane by a few tens of \kms. We discuss possible explanations for this
  observation, including outflowing gas in a multiphase fountain flow
  and ram pressure effects from the interaction with NGC 5774.
\item This and previous works indicate the need for a supplemental source of
  heating and/or ionization that behaves differently from photoionization in
  the eDIG layer of NGC 5775. The extended, co-spatial thermal and nonthermal
  halos in this galaxy, as well as evidence that cosmic-ray heating may be
  important \citep{Wiener2013}, motivate further study of the role of cosmic
  rays in the energy balance and ionization state of the eDIG.
\end{itemize}

As discussed in \S\ref{chap3:sec:disc}, past and current observations of the
eDIG layer in NGC 5775 are most consistent with a non-hydrostatic, galactic
fountain flow. Turbulent, multiphase gas clouds ejected from the disk reach
scale heights characteristic of their ejection energies, possibly passing
through phase transitions that obscure some stages of this process in the
optical. Shock heating and/or ionization evidenced by trends in emission-line
ratios are consistent with the interaction of ejected clouds with the ambient
medium. In this picture, the gas in the eDIG layer both originates in and
returns to the disk and thus is not indicative of a substantial transfer of
baryons or metals between the disk and the environment. Enriching and cooling
of hot halo gas as cool clouds travel through the lower halo may, however,
contribute to the presence of warm, ionized gas at the disk-halo interface.

To further test this interpretation, we suggest future observations to shed
light on several open questions about the eDIG layer in this and similar
systems. In NGC 5775, the remarkable vertical extent of the gaseous halo in
emission is intriguing. It is not yet clear that we have detected the full
vertical extent in emission, nor have we definitely determined the number of
exponential electron density components. Deep, ground-based, narrow-band
imaging that prioritizes both the field of view and quality sky subtraction
would help to achieve these goals, furthering our understanding of the
fraction of halo gas mass found in the warm, ionized phase.

Additionally, interpretation of intriguing kinematic results such as the
blueshifted gas at high $z$ along the minor axis is challenging with the
limited spatial coverage of longslit spectroscopy. Integral field unit and/or
Fabry-Perot spectroscopy would enable analysis of gas kinematics over greatly
improved fields of view, an effort begun for this galaxy by
\citet{Heald2006a}. We propose additional observations with improved
sensitivity in order to probe the kinematics at $|z| \ge 4$ kpc in a
spatially-resolved fashion. Such observations would enable discrimination
between outflow and ram pressure origins for the minor axis kinematics as
discussed in \S\ref{chap3:sec:disc}. In the modeling regime, improved models
of cosmic-ray transport will advance our understanding of the validity of the
energy equipartition assumption in assessing nonthermal pressure gradients in
gaseous halos \citep[e.g.,][]{Schmidt2016}.

Finally, it is clear that edge-on galaxies alone will not fully illuminate the
vertical structure, support, and kinematics of eDIG layers. Ongoing efforts to
study the vertical bulk velocities and velocity dispersions in low-inclination
galaxies will be a powerful complement to the high-inclination perspective
\citep[e.g.,][]{Boettcher2017}, allowing more rigorous testing of the ideas
discussed here.

\acknowledgments{All of the observations reported in this paper were obtained
  with the Robert Stobie Spectrograph (RSS) on the Southern African Large
  Telescope (SALT) under programs 2015-1-SCI-023 and 2016-2-SCI-029 (PI:
  E. Boettcher). We thank the SALT astronomers and telescope operators for
  obtaining the observations and Petri Vaisanen for advice on data
  acquisition. We acknowledge Ken Nordsieck for his expertise on the RSS
  instrument and Arthur Eigenbrot for his help with data reduction. We thank
  Bob Benjamin, Rainer Beck, George Heald, Dick Henriksen, Judith Irwin,
  Jiang-Tao Li, Richard Rand, Carlos Vargas, and Rene Walterbos for useful
  discussions and comments. We acknowledge Masataka Okabe and Kei Ito for
  supplying the colorblind-friendly color palette used in this paper (see
  \url{fly.iam.u-tokyo.ac.jp/color/index.html}).

  This material is based upon work supported by the National Science
  Foundation Graduate Research Fellowship Program under Grant
  No. DGE-1256259. Any opinions, findings, and conclusions or recommendations
  expressed in this material are those of the author(s) and do not necessarily
  reflect the views of the National Science Foundation. Support was also
  provided by the Graduate School and the Office of the Vice Chancellor for
  Research and Graduate Education at the University of Wisconsin-Madison with
  funding from the Wisconsin Alumni Research Foundation.

  This work has made use of NASA's Astrophysics Data System and of the
  NASA/IPAC Extragalactic Database (NED) which is operated by the Jet
  Propulsion Laboratory, California Institute of Technology, under contract
  with the National Aeronautics and Space Administration.}

\section{Appendix}
\label{chap3:sec:appendix}

\subsection{Example Spectra and Gaussian Fits}
\label{chap3:sec:ex}

\begin{figure*}[h]
  \epsscale{1.0}\plotone{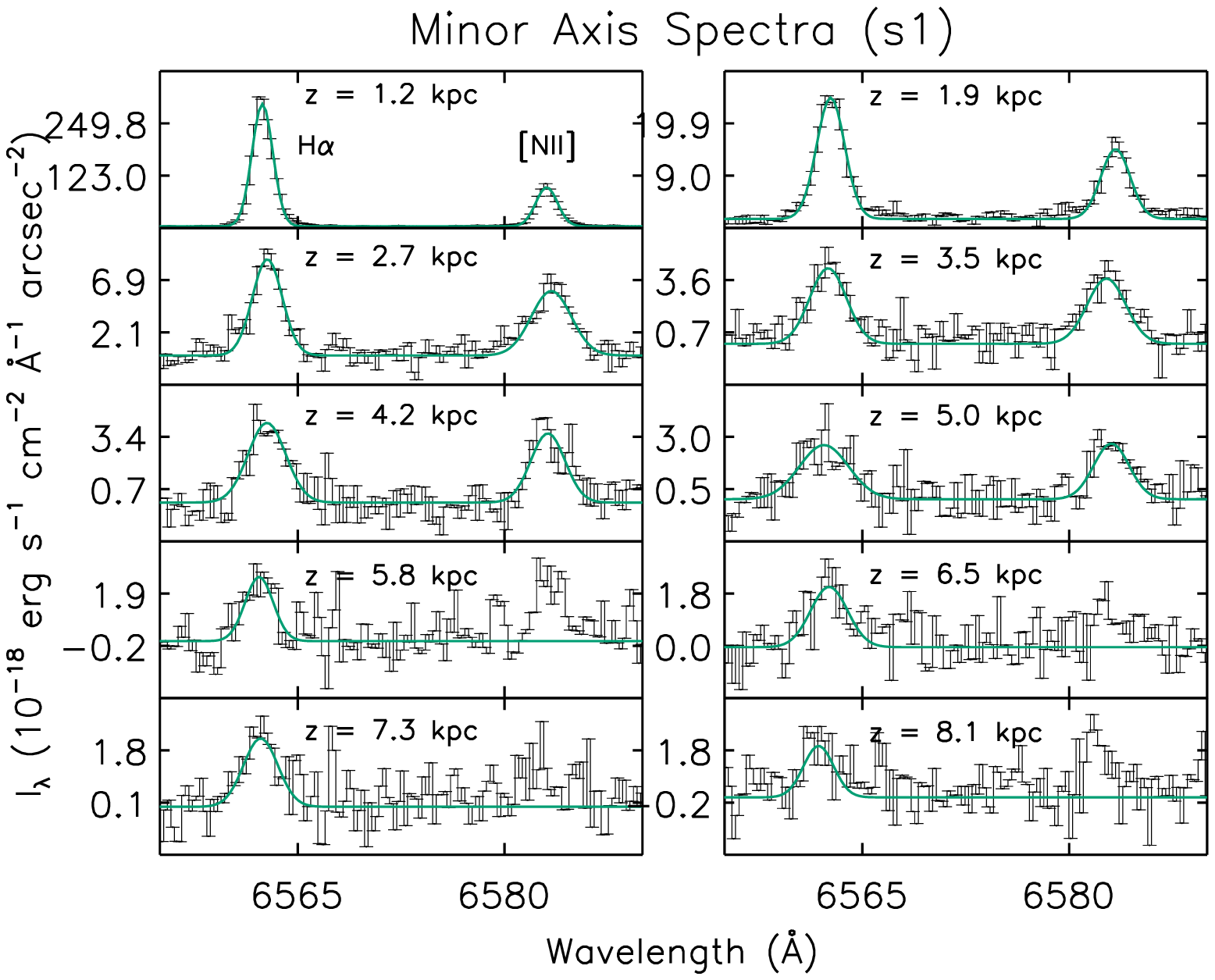}
  \epsscale{1.0}\plotone{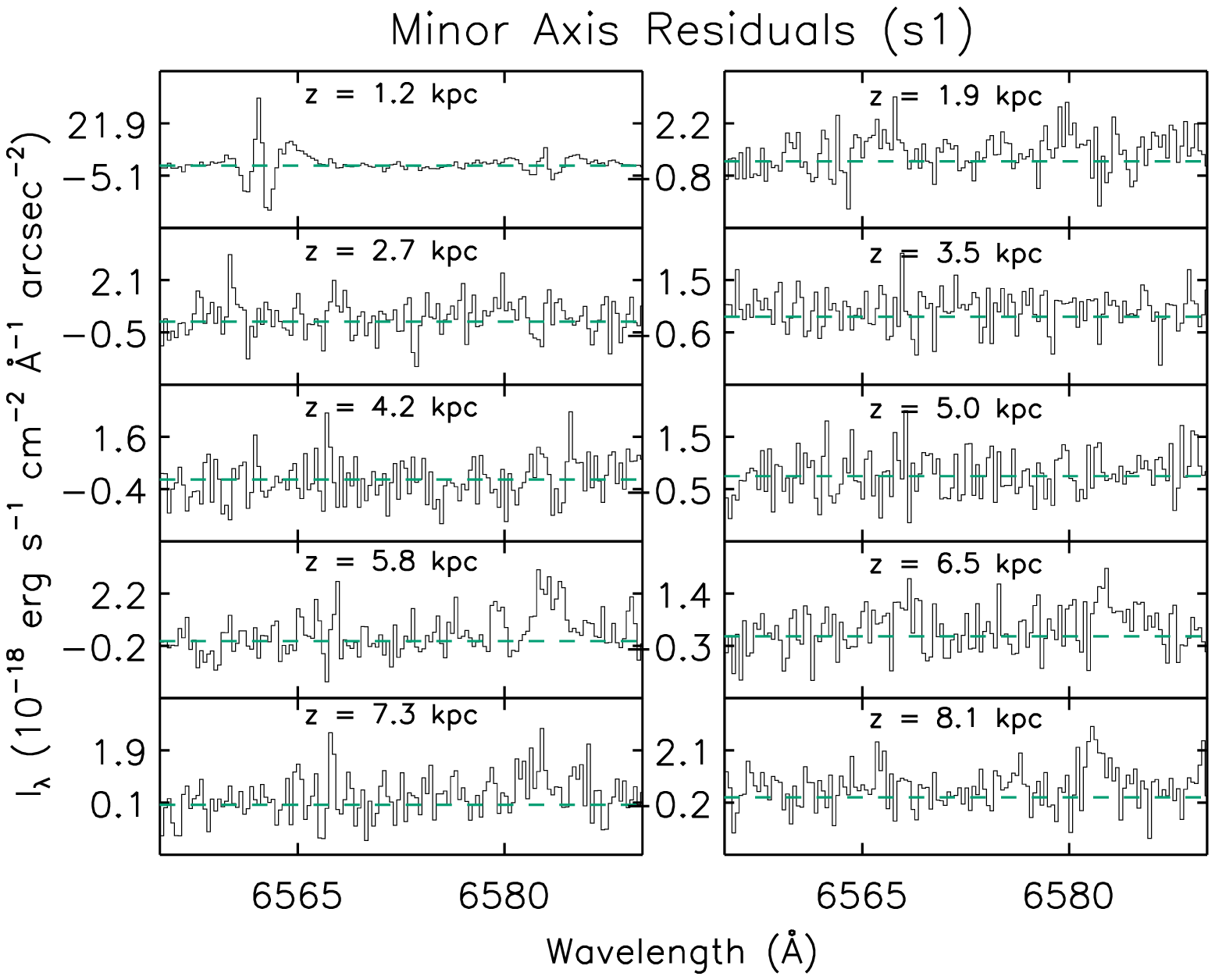}
  \caption{Example rest-frame spectra along the minor axis on the northeast
    side of the galaxy (s1). Gaussian fits and their residuals are shown in
    the top and bottom panels, respectively. Fits are shown for the H$\alpha$
    and [NII]$\lambda$6583 lines that meet our detection criteria.}
  \label{chap3:fig11}
\end{figure*}

In Fig. \ref{chap3:fig11}, we show example spectra from the minor axis on the
northeast side of the galaxy (s1). In the top panel, the Gaussian fits for
detected H$\alpha$ and [NII]$\lambda$6583 lines are shown; the corresponding
residuals are given in the bottom panel. In general, the emission-line
profiles are well characterized by a single Gaussian fit. However, at high
S/N, low- and/or high-velocity wings are apparent in some locations. This is
particularly evident for the red wing seen in both the H$\alpha$ and [NII]
residuals at $z = 1.2$ kpc. While detection of such wings is challenging at
high $|z|$ due to the low S/N, the detection of these wings at low $|z|$
suggests the presence of gas at more extreme velocities than are considered in
this work. Higher S/N spectra that can better characterize the emission-line
wings would help to determine whether there is an additional,
low-surface-brightness component with a higher velocity dispersion than
quantified here.

We provide the best-fit Gaussian parameters in machine-readable tables
available online (see Tab. \ref{chap3:tab3} - \ref{chap3:tab6}). Portions of
these tables are given below for guidance with respect to their form and
content.

\begin{deluxetable*}{cccccccc}
\tabletypesize{\scriptsize}
\tablecolumns{8}
\tablewidth{0pt}
\tablecaption{Best-fit Emission-line Properties: s1}
\tablehead{ 
\colhead{$z$} &
\colhead{$I_{H\alpha}$} &
\colhead{$v_{H\alpha}$\tablenotemark{a}} &
\colhead{$\sigma_{H\alpha}$\tablenotemark{b}} &
\colhead{$\frac{[OI]\lambda6300}{H\alpha}$} &
\colhead{$v_{[OI]}$} &
\colhead{$\sigma_{[OI]}$} &
\colhead{...\tablenotemark{c}}\\
\colhead{(kpc)} &
\colhead{($10^{-17}$ erg cm$^{-2}$ s$^{-1}$ arcsec$^{-2}$)} &
\colhead{(km s$^{-1}$)} &
\colhead{(km s$^{-1}$)} &
\colhead{} &
\colhead{(km s$^{-1}$)} &
\colhead{(km s$^{-1}$)} &
\colhead{}
}
\startdata
8.1 & 0.5 $\pm$ 0.1 & 1637 $\pm$ 7 & 42 $\pm$ 7 & 0 & 0 & 0 & ... \\
... & ... & ... & ... & ... & ... & ... & ...
\enddata
\tablenotetext{a}{The heliocentric radial velocity.}  \tablenotetext{b}{The
  line width corrected for the spectral resolution. $\sigma$ refers to the
  standard deviation and not the FWHM of the Gaussian fit.}
\tablenotetext{c}{The remaining columns give the line ratio with respect to
  H$\alpha$, velocity, and velocity dispersion for the [NII]$\lambda$6583 and
  [SII]$\lambda$6717 lines. Null entries indicate locations where the relevant
  emission line was not detected. This table is available in its entirety in
  machine-readable form.}
\label{chap3:tab3}
\end{deluxetable*}

\begin{deluxetable*}{cccccccc}
\tabletypesize{\scriptsize}
\tablecolumns{8}
\tablewidth{0pt}
\tablecaption{Best-fit Emission-line Properties: s2}
\tablehead{ 
\colhead{$R$} &
\colhead{$I_{H\alpha}$} &
\colhead{$v_{H\alpha}$\tablenotemark{a}} &
\colhead{$\sigma_{H\alpha}$\tablenotemark{b}} &
\colhead{$\frac{[NII]\lambda6583}{H\alpha}$} &
\colhead{$v_{[NII]}$} &
\colhead{$\sigma_{[NII]}$} &
\colhead{...\tablenotemark{c}}\\
\colhead{(kpc)} &
\colhead{($10^{-17}$ erg cm$^{-2}$ s$^{-1}$ arcsec$^{-2}$)} &
\colhead{(km s$^{-1}$)} &
\colhead{(km s$^{-1}$)} &
\colhead{} &
\colhead{(km s$^{-1}$)} &
\colhead{(km s$^{-1}$)} &
\colhead{}
}
\startdata
13.5 & 1.1 $\pm$ 0.1 & 1532 $\pm$ 5 & 46 $\pm$ 5 & 0 & 0 & 0 & ... \\
... & ... & ... & ... & ... & ... & ... & ...
\enddata
\tablenotetext{a}{The heliocentric radial velocity.}  \tablenotetext{b}{The
  line width corrected for the spectral resolution. $\sigma$ refers to the
  standard deviation and not the FWHM of the Gaussian fit.}
\tablenotetext{c}{The remaining columns give the line ratio with respect to
  H$\alpha$, velocity, and velocity dispersion for the [SII]$\lambda$6717
  line. Null entries indicate locations where the relevant emission line was
  not detected. This table is available in its entirety in machine-readable
  form.}
\label{chap3:tab4}
\end{deluxetable*}

\begin{deluxetable*}{cccccccc}
\tabletypesize{\scriptsize}
\tablecolumns{8}
\tablewidth{0pt}
\tablecaption{Best-fit Emission-line Properties: s3uv}
\tablehead{ 
\colhead{$z$} &
\colhead{$I_{[OII]}$} &
\colhead{$\frac{[OII]\lambda 3729}{[OII]\lambda 3726}$} &
\colhead{$v_{[OII]}$\tablenotemark{a}} &
\colhead{$\sigma_{[OII]}$\tablenotemark{b}}\\
\colhead{(kpc)} &
\colhead{(SALT ADU)} &
\colhead{} &
\colhead{(km s$^{-1}$)} &
\colhead{(km s$^{-1}$)} 
}
\startdata
5.8 & 11 $\pm$ 9 & 2.0$^{+0.83}_{-0.83}$ & 1651$^{+16}_{-16}$ & 30$^{+26}_{-16}$ \\
... & ... & ... & ... & ... & 
\enddata
\tablenotetext{a}{The heliocentric radial velocity.}  \tablenotetext{b}{The
  line width corrected for the spectral resolution. $\sigma$ refers to the
  standard deviation and not the FWHM of the Gaussian fit. This table is
  available in its entirety in machine-readable form.}
\label{chap3:tab5}
\end{deluxetable*}

\begin{deluxetable*}{cccccccc}
\tabletypesize{\scriptsize}
\tablecolumns{8}
\tablewidth{0pt}
\tablecaption{Best-fit Emission-line Properties: s3o}
\tablehead{ 
\colhead{$z$} &
\colhead{$I_{H\alpha}$} &
\colhead{$v_{H\alpha}$\tablenotemark{a}} &
\colhead{$\frac{[OII]\lambda3727}{H\alpha}$} &
\colhead{$\frac{[OIII]\lambda5007}{H\alpha}$} &
\colhead{$\frac{[OI]\lambda6300}{H\alpha}$} &
\colhead{$\frac{[NII]\lambda6583}{H\alpha}$}\\
\colhead{(kpc)} &
\colhead{(SALT ADU)} &
\colhead{} &
\colhead{} &
\colhead{} &
\colhead{} &
\colhead{}
}
\startdata
8.1 & 96 $\pm$ 17 & 1714 $\pm$ 23 & 0 & 0 & 0 & 0 \\
... & ... & ... & ... & ... & ... & ...
\enddata
\tablenotetext{a}{The heliocentric radial velocity. Null entries indicate
  locations where the relevant emission line was not detected. This table is
  available in its entirety in machine-readable form.}
\label{chap3:tab6}
\end{deluxetable*}

\subsection{A Mass Model for NGC 5775}
\label{chap3:sec:mass_mod}

\begin{figure}[h]
  \epsscale{1.2}\plotone{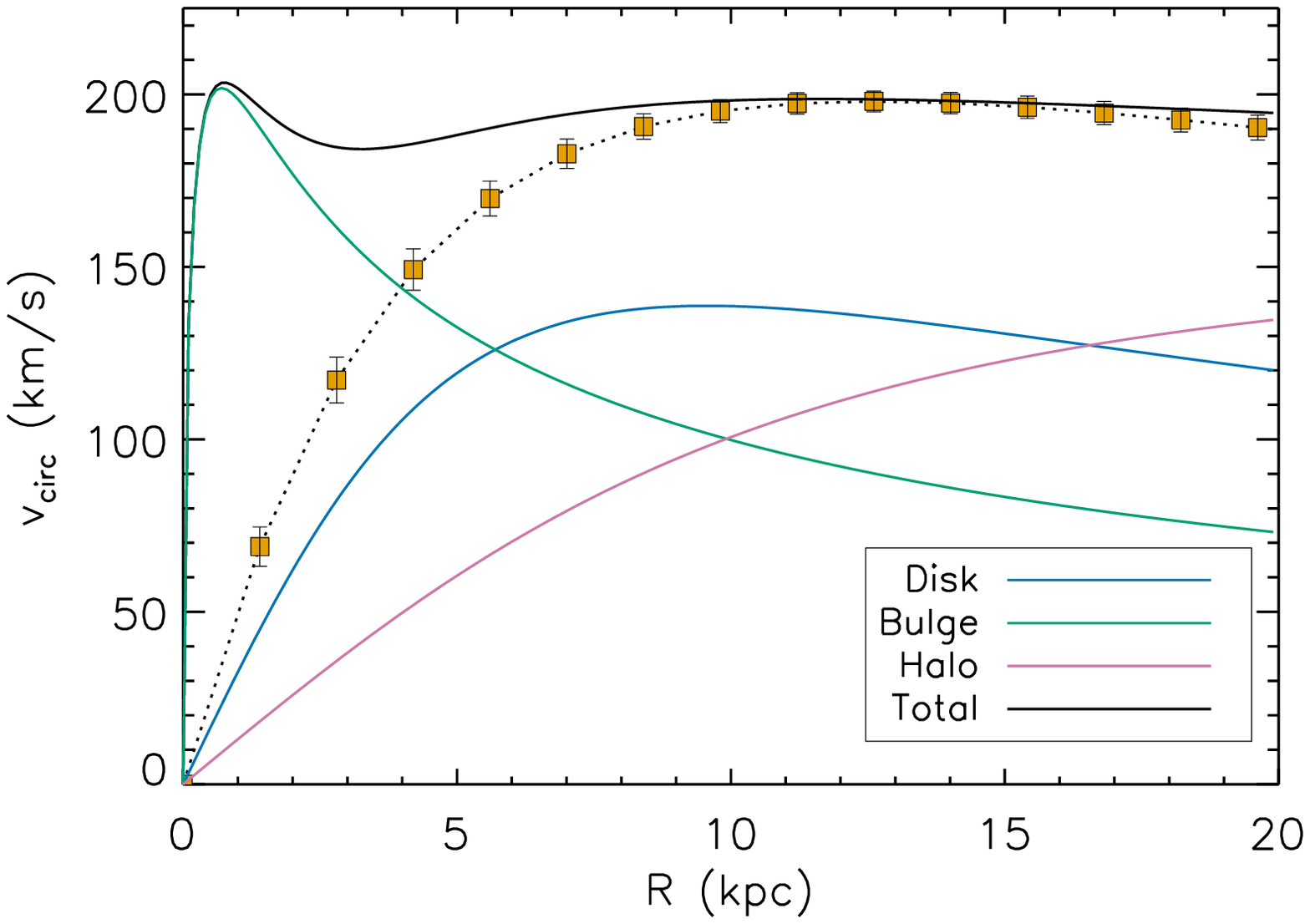}
  \caption{Galactic rotation curve of \citet{Wolfire1995} adapted to NGC 5775
    (solid lines) compared to the modeled HI rotation curve of
    \citet{Irwin1994} (yellow squares).}
  \label{chap3:fig12}
\end{figure}

To determine the galactic gravitational potential of NGC 5775, we follow
\citet{Collins2002} in adopting the mass model presented in the Appendix of
\citet{Wolfire1995}. This model consists of a disk, spherical bulge, and
logarithmic halo; although it is developed for the Galaxy, it can be adapted
to other systems by adjusting the $v_{circ}$ parameter. We take $v_{circ} =
195$ \kms to match the HI rotation curve of NGC 5775 \citep{Irwin1994}, and we
compare the adapted \citet{Wolfire1995} and \citet{Irwin1994} rotation curves
in Fig. \ref{chap3:fig12}. The rotation curves are in good argeement at large
$R$; discrepancy at small $R$ may be due to lack of HI in the inner part of
the galaxy. For the disk, bulge, and halo components, the gravitational
potential, $\Phi(z,R)$, and the gravitational acceleration, $-g_{z} =
-\frac{d\Phi(z,R)}{dz}$, are given in Eq. (A1) - (A3) and (A7) - (A9) of
\citet{Wolfire1995}.

\bibliographystyle{apj}

\end{document}